%% file: final.tex
\documentclass[fleqn,usenatbib]{mnras}

\usepackage{newtxtext,newtxmath}

\usepackage[T1]{fontenc}

\DeclareRobustCommand{\VAN}[3]{#2}
\let\VANthebibliography\thebibliography
\def\thebibliography{\DeclareRobustCommand{\VAN}[3]{##3}\VANthebibliography}

\usepackage{graphicx}	
\usepackage{amsmath}	
\usepackage{xspace}     
\usepackage{natbib}
\usepackage{xcolor}
\usepackage{framed}

\usepackage{longtable}
\usepackage{makecell}
\usepackage{xtab,booktabs}
\usepackage{soul}

\newcommand{\chandra}{\textit{Chandra}\xspace}
\newcommand{\xrism}{\textit{XRISM}\xspace}
\newcommand{\Rc}[1]{$R_{\rm #1 c}$}

\newcommand{\drrk}{$\left(\delta \rho / \rho \right)_{k}$}
\newcommand{\drr}{$\delta \rho / \rho$}

\definecolor{shadecolor}{gray}{0.97}

\title[Merger-driven ICM density perturbations]{Merger-driven multi-scale ICM density perturbations: testing cosmological simulations and constraining plasma physics}

\author[A. Heinrich et al.]{
Annie Heinrich,$^{1}$\thanks{E-mail: amheinrich@uchicago.edu}
Irina Zhuravleva,$^{1}$
Congyao Zhang,$^{1}$
Eugene Churazov,$^{2,3}$
William Forman,$^{4}$
\newauthor \
Reinout~J. van Weeren$^{5}$
\\
$^{1}$Department of Astronomy \& Astrophysics, University of Chicago, 5640 S Ellis Avenue, Chicago, IL 60637, USA\\
$^{2}$Max Planck Institute for Astrophysics, Karl-Schwwarzschild-Strasse 1, D-85478 Garching, Germany\\
$^{3}$Space Research Institute (IKI), Profsoyuznaya 84/32, Moscow 117997, Russia\\
$^{4}$Center for Astrophysics, Harvard \& Smithsonian, 60 Garden St., MS-3, Cambridge, MA 02138, USA\\
$^{5}$Leiden Observatory, Leiden University, PO Box 9513, 2300 RA Leiden, the Netherlands\\
}

\date{Accepted 2024 January 12. Received 2024 January 10; in original form 2023 November 15}

\pubyear{2024}

\begin{document}
\label{firstpage}
\pagerange{\pageref{firstpage}--\pageref{lastpage}}
\maketitle

\begin{abstract}

The hot intracluster medium (ICM) provides a unique laboratory to test multi-scale physics in numerical simulations and probe plasma physics.
Utilizing archival \chandra observations, we measure density fluctuations in the ICM in a sample of 80 nearby ($z \lesssim 1$) galaxy clusters and infer scale-dependent velocities within regions affected by mergers ($r < $ \Rc{2500}), excluding cool-cores. 
Systematic uncertainties (e.g., substructures, cluster asymmetries) are carefully explored to ensure robust measurements within the bulk ICM. 
We find typical velocities $\sim 220$ ($300$) km s$^{-1}$ in relaxed (unrelaxed) clusters, which translate to non-thermal pressure fractions $\sim 4$ ($8$) per cent, and clumping factors $\sim 1.03$ ($1.06$). 
{We show that density fluctuation amplitudes could distinguish relaxed from unrelaxed clusters in these regions.}
Comparison with density fluctuations in cosmological simulations shows good agreement in merging clusters.
Simulations underpredict the amplitude of fluctuations in relaxed clusters on length scales $< 0.75$ \Rc{2500}, suggesting these systems are most sensitive to ``missing'' physics in the simulations.
In clusters hosting radio halos, we examine correlations between gas velocities, turbulent dissipation rate, and radio emission strength/efficiency to test turbulent re-acceleration of cosmic ray electrons. 
We measure a weak correlation, driven by a few outlier clusters, in contrast to some previous studies. 
Finally, we present upper limits on effective viscosity in the bulk ICM of 16 clusters, showing it is systematically suppressed by at least a factor of 8, and the suppression is a general property of the ICM. 
Confirmation of our results with direct velocity measurements will be possible soon with \xrism.

\end{abstract}

\section{Introduction}
\label{sec:introduction} 
Galaxy clusters are the most massive gravitationally-bound systems in the universe and continue to grow via hierarchical mergers and continuous accretion of matter from the intergalactic medium.
These accretion processes are the dominant drivers of gas dynamics within the hot ($T\sim 10^{7.5}$ K) intracluster medium (ICM) outside the innermost regions \citep[e.g.,][]{walker_physics_2019, heinrich_constraining_2021, zuhone_merger_2022} that are often affected by feedback from central supermassive black holes \citep[see, e.g.,][for a recent review]{hlavacek-larrondo_agn_2022}.
Probing turbulent and bulk motions within the ICM is crucial for understanding energy partition during large-scale structure evolution, astrophysical processes that drive the evolution of galaxies within the ICM, and plasma physics, yet direct measurements of these motions remain challenging \citep[for exceptions, see, e.g.,][]{gatuzz_measuring_2023, sanders_measuring_2020,hitomi_collaboration_atmospheric_2018, sanders_velocity_2013}. 
Directly resolving bulk and turbulent velocities of the hot gas in galaxy clusters, groups, and massive galaxies will soon be possible with the recently-launched \xrism  satellite \citep{xrism_science_team_science_2020}.

While waiting for data from X-ray microcalorimeters, such as \xrism, \textit{NewAthena} \citep{barret_athena_2020} and \textit{LEM} \citep{kraft_line_2022}, to become available, indirect techniques have been used to constrain ICM velocities, including resonant scattering \citep[e.g.,][]{hitomi_collaboration_measurements_2017, ogorzalek_improved_2017}, inference from metal abundance distributions \citep{rebusco_impact_2005}, shock widths \citep{nulsen_deep_2013}, and X-ray surface brightness or Sunyaev-Zel'dovich fluctuations \citep[e.g.,][]{schuecker_probing_2004, churazov_x-ray_2012, zhuravleva_gas_2015,walker_constraining_2015, romero_inferences_2023}.
This last method is particularly attractive as it provides characteristic velocity amplitudes as a function of spatial scale (or wavenumber) --- information that is difficult to extract, even from direct velocity measurements through Doppler shifts.

The X-ray surface brightness method utilizes a relation between the scale-dependent amplitude (related to a power spectrum) of gas density fluctuations, \drrk{}, and the one-dimensional characteristic velocity, $V_\text{1D,k}$. 
This method has previously been calibrated using relaxed clusters from non-radiative cosmological simulations \citep{zhuravleva_relation_2014} and idealized Coma-like cluster simulations \citep{gaspari_relation_2014}. 
It was first applied to a sample of the brightest cores of relaxed clusters to measure velocity power spectra \citep[e.g.,][]{zhuravleva_turbulent_2014,zhuravleva_gas_2018}.
Recently, this relationship was verified for unrelaxed objects in cosmological simulations \citep{ simonte_exploring_2022,zhuravleva_indirect_2023}, showing that the correlation is still valid, though the scatter is larger by a factor of $\sim 2$ compared to relaxed systems \citep{zhuravleva_indirect_2023}. 
High-resolution idealized simulations of stratified turbulence \citep{mohapatra_turbulence_2020,mohapatra_turbulent_2021} showed that the relationship is more complicated and depends on the level of stratification, turbulence driving, and pressure/entropy scale height of the system, which could explain the scatter in the simplified relation.

Using mostly imaging data from \textit{XMM-Newton}, the technique has been applied to unrelaxed objects and regions outside of cool-cores \citep{eckert_connection_2017, zhang_planck_2022}, where gas dynamics are driven mainly by cluster mergers.
The relatively large PSF of that telescope, however, limits measurements to rather large scales.
Given high-resolution images of galaxy clusters obtained with \chandra, this method can provide velocity measurements on scales small enough to test plasma effects and small-scale physics in simulations. Luckily, after nearly 25 years of operations, the \chandra archive contains enough data to measure these velocities in a large sample of clusters.

Velocity measurements outside of cool-cores open up a variety of phenomena that can be studied.
For example, turbulence in the ICM is theorized to be linked to diffuse synchrotron emission from giant radio halos, mostly found in dynamically disturbed clusters \citep[e.g.,][]{cassano_connection_2010}.
The GeV cosmic-ray electrons (CRe) powering the radio emission in halos are likely associated with 2nd-order Fermi acceleration \citep[e.g.,][]{brunetti_compressible_2007, petrosian_particle_2008, van_weeren_diffuse_2019} of a seed CRe population. Measurements of turbulent velocities and their correlation with the radio emission strength could, therefore, provide valuable insights into the CRe acceleration mechanisms.

Additionally, the ICM is threaded by a (likely) tangled magnetic field with $B\sim$ a few $\mu G$.
Although the magnetic pressure is small compared to the thermal pressure, the field can significantly alter the ICM microphysics, including the ICM's transport properties \citep[e.g.,][]{schekochihin_turbulence_2006,kunz_plasma_2022,arzamasskiy_kinetic_2023}.
Multi-scale velocity measurements {to scales close to the Coulomb mean-free-path of particle interactions ($\lambda_\text{mfp}$)} are crucial for understanding the ICM as a weakly-collisional high-$\beta$ plasma.
{Since $\lambda_\text{mfp}$ typically increases with the radius in the ICM, probing microphysical scales with observations is most promising in regions outside dense cluster cores. 
}

{Finally, the partition of energy during matter accretion onto virialized halos and cluster mergers also depends on the poorly known microphysics of the hot plasma in the ICM. 
Measuring gas motions outside cluster cores, one can directly probe what fraction of gravitational energy converts into thermal and kinetic energies. 
The latter contribution also biases cluster mass measurements based on the assumption of hydrostatic equilibrium \citep[e.g.,][]{lau_residual_2009, nelson_weighing_2014, vazza_turbulent_2018}. Measuring the non-thermal pressure and demonstrating that it is responsible for the mass bias are important for a few percent-level precision cluster cosmology {with X-ray and SZ techniques.}}

In this work, we will, {for the first time,} measure ICM gas velocities across a range of scales using the density fluctuations technique in a large sample of galaxy clusters within a cosmologically-interesting region of radius of \Rc{2500} using archival \chandra data. 
The high spatial resolution of \chandra allows us to measure velocities on smaller scales (and a broader range of scales) than studied previously.
We will probe gas dynamics driven by mergers and large-scale structure evolution, minimizing the effects of AGN feedback and gas cooling.
We assume a flat $\Lambda$CDM cosmology with $h=0.7$ and $\Omega_{\rm m}=0.3$. 
$R_{n\text{c}}$ is defined as the cluster-centric radius within which the average density is $n$ times the critical density of the universe at the cluster's redshift.

In Section \ref{sec:methods}, we describe our sample of clusters, image processing techniques, and power spectral analysis methods.
In Section \ref{sec:results}, we present individual and averaged measurements of \drrk{}, the characteristic ICM gas velocity $V_\text{1D,k}$, and place limits on the ICM clumping factor {and scale-independent characteristic velocity}.
Section \ref{sec:simulations} discusses these constraints in relation to cosmological simulations.
Finally, Section \ref{sec:plasma} focuses on relevant plasma physics, namely, testing the turbulent re-acceleration mechanism of CRe and constraining transport properties in the bulk ICM.

\section{Sample selection and data analysis}
\label{sec:methods}

\begin{figure}
\vspace{-0.4cm}
    \centering
    \includegraphics[trim=0 15 0 -20, width=\linewidth]{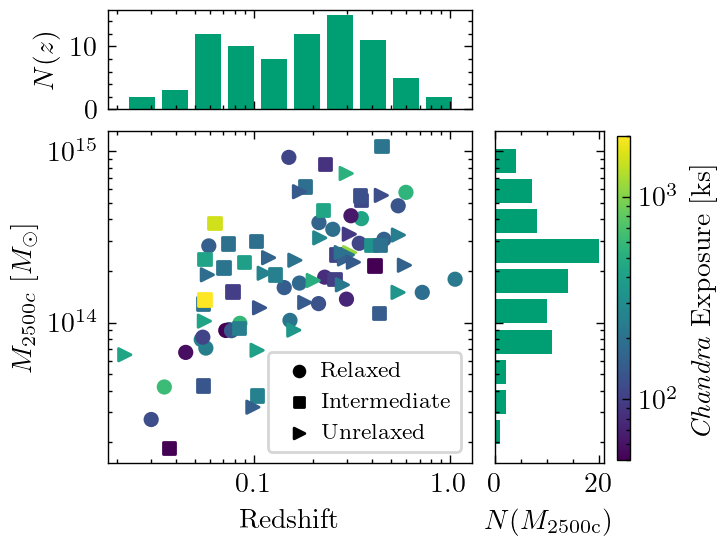}
    \caption{Our sample of 80 galaxy clusters in which we measure density fluctuation power spectra: $M_{2500c}$ vs. cluster redshift. Marker shapes indicate cluster morphology: ``relaxed'' (circles, cool-core systems with no signs of recent mergers), ``intermediate'' (squares, evidence of a minor merger), or ``unrelaxed'' (triangles, recent or ongoing major merger, no cool-cores), based on a literature review and visual inspection of the X-ray images, see Section \ref{sec:classification} for details. Color indicates flare-removed (clean) \chandra exposure time in ks. \textbf{Top}: histogram of cluster redshifts. \textbf{Right}: histogram of cluster masses within \Rc{2500}.}
    \label{fig:sample}
\end{figure}

We aim to measure the power spectra of hot gas density fluctuations down to the smallest possible scales in a large sample of galaxy clusters. 
The \chandra X-ray Observatory is best-suited for our goal as it offers the highest angular resolution.
We select systems where archival \chandra observations cover (nearly) the entirety of the cluster within a projected radius of \Rc{2500} from the cluster center, as determined from X-ray surface brightness contours at the same radius.
The clusters are also brighter than the background level by at least a factor of two  in this region.
To have a sufficient number of photons in the X-ray images, we further require the total exposure time of each cluster to be greater than 100 ks.
However, we also examine a few bright clusters below this threshold.
Our final sample totals 80 clusters, with \chandra ObsIDs and cleaned exposure times presented in Table \ref{tab:observations}. 
We collected measured total masses ($M_{\rm 2500c}$) and characteristic radii ($R_{\rm 2500c}$) from the literature\footnote{Where measurements of \Rc{2500} were unavailable, we assume $R_\text{2500c}=0.4 R_\text{500c}$ \citep{arnaud_universal_2010}.}, summarizing their values and corresponding references in Table \ref{tab:parameters}. 
Figure \ref{fig:sample} summarizes our sample. 
Most clusters in our sample have masses ($M_{\rm 2500c}$) between $\sim 8 \times 10^{13}$ and $\sim 10^{15}$ $M_\odot$, and have redshifts between $\sim 0.05$ and $\sim 0.5$.

\begin{figure*}
    \centering
    \includegraphics[trim=0 15 0 0, width=\textwidth]{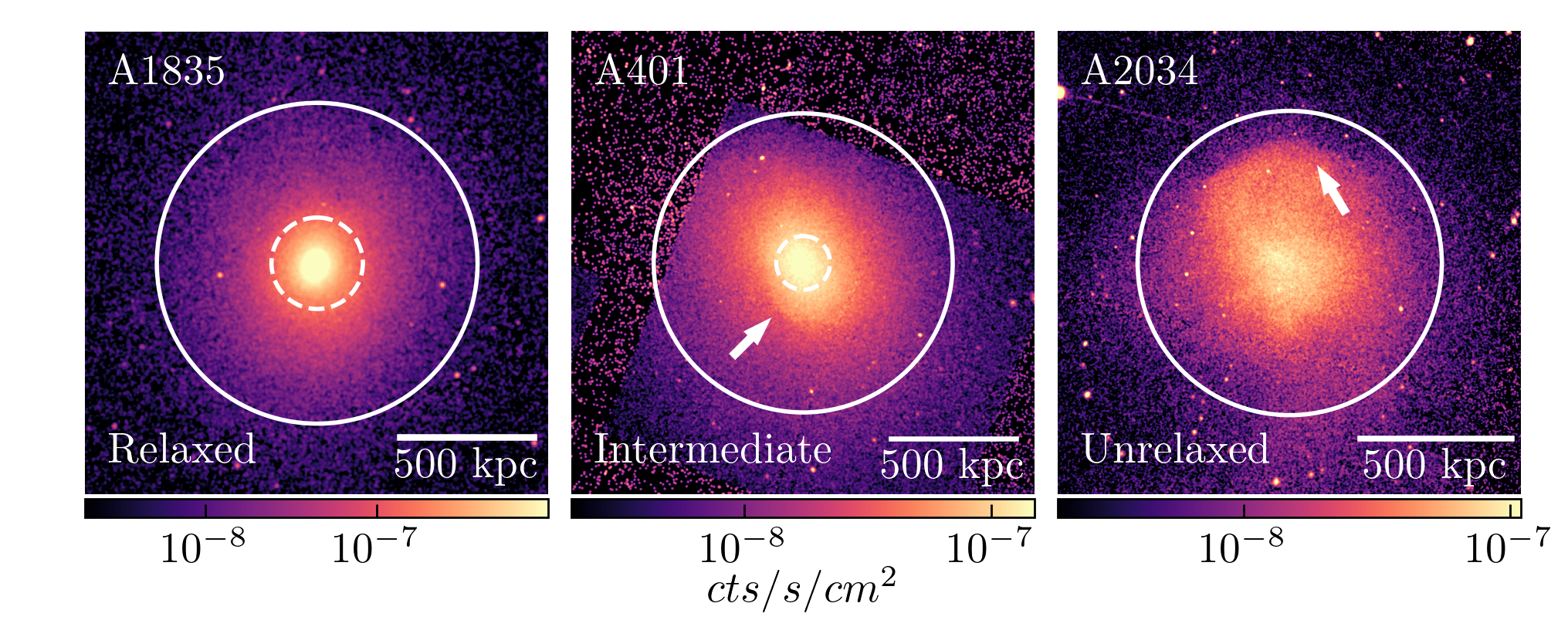}
    \caption{Representative examples of clusters from relaxed, intermediate, and unrelaxed subsamples. Exposure-corrected \chandra X-ray (0.5-3.5 keV band) images  are shown, lightly smoothed with a Gaussian for display purposes. Solid white circles represent \Rc{2500}, while dashed circles mark the cool-core radius if the cluster has one. \textbf{A1835} is a relaxed, cool-core cluster that does not show any signs of mergers \citep{Schmidt2001,Govoni2009}. \textbf{A401} is a cool-core cluster, however, it is undergoing a merger as evidenced by the sloshing cold front to the south (white arrow), the proximity of the cluster to A399, and the presence of a giant radio halo \citep{murgia_double_2010}. For these reasons, we place it in the intermediate category. \textbf{A2034} is a clear example of an unrelaxed, merging cluster as it lacks a cool-core, shows irregular morphology, contains a merger shock north of the cluster center \citep[white arrow][]{owers_merger_2013}, and hosts a radio halo \citep{botteon_planck_2022}.} 
    \label{fig:morphologies}
\end{figure*}

\subsection{Initial data processing}
\label{sec:observations}
The data for each cluster are reprocessed using the standard methods described in \citet{vikhlinin_chandra_2005}, implementing scripts from \textsc{ciao} version 4.14. 
Event lists are reprocessed using CalDB version 4.9.7.
Background flares are removed by filtering lightcurves using the \textsc{lc\_sigma\_clip} script. 
Non-X-Ray background files are created using combined products of the \textsc{blanksky} and \textsc{readout\_bkg} routines.
Event lists are restricted to the 0.5-3.5 keV band, where X-ray emissivity is almost independent of temperature for typical temperatures of galaxy clusters \citep[see their Figure 1]{forman_filaments_2007}.
The resultant event lists are combined using the \textsc{merge\_obs} script, creating mosaic images of X-ray counts, background, a weighted instrument exposure map, and an exposure-weighted point-spread-function (PSF) map. 
A selection of exposure-corrected images of galaxy clusters from our sample is presented in Figure \ref{fig:morphologies}.

\subsection{Classification of galaxy clusters}
\label{sec:classification}

Based on these X-ray images and a literature review, we classify clusters in our sample into ``relaxed'', ``intermediate'', and ``unrelaxed'' systems.
Relaxed clusters have cool-cores\footnote{{Here, cool-core {regions} are defined as central {regions with} cooling times less than the Hubble time.}}, exhibit regular spherical or elliptical morphologies in X-rays, and show no signs of recent or ongoing mergers. 
Most intermediate clusters maintained their cool-cores and showed evidence of minor mergers (e.g., gas sloshing). 
Clusters that are undergoing a merger but show regular morphology (e.g., A1689, which has spherical X-ray morphology and a weak cool-core, but hosts a radio halo \citep{vacca_discovery_2011} and is likely undergoing a merger along the line-of-sight based on the galaxy redshift distribution \citep{girardi_new_1997}) are also included into the ``intermediate'' category.
Unrelaxed clusters are generally non-cool-core and show clear signs of ongoing major mergers, including shock fronts and ``bullet-like'' morphologies.
Representative examples of these three categories are shown in Figure \ref{fig:morphologies}. 
While the divisions between relaxed and intermediate clusters, and between intermediate and unrelaxed clusters, are somewhat uncertain, the relaxed and unrelaxed subsamples are likely well separated in terms of merger effects.
The ``relaxed'', ``intermediate'', and ``unrelaxed'' subsamples include 24, 30, and 26 clusters, respectively.

\subsection{Deprojection analysis}

To obtain radial profiles of gas density, temperature, and abundance of heavy elements for a subsample of galaxy clusters discussed in Section \ref{sec:viscosity}, we extract projected spectra in a set of radial annuli. 
These spectra are deprojected using a flavor of the onion-peeling technique described by \citet{churazov_xmm-newton_2003}. 
The resulting deprojected spectra are fit in the broad $0.6-8$ keV band using \textsc{XSPEC} code \citep{arnaud_xspec_1996} and \textsc{APEC} plasma model \citep{smith_collisional_2001} based on \textsc{AtomDB} \citep{foster_updated_2012} version 3.0.9. 
The abundance of heavy elements is fixed at 0.5 relative to Solar abundances of heavy elements taken from \citet{anders1989} in the fitting. 
However, we verify that treating the abundance as a free parameter does not affect the measured density and temperature profiles.  

\subsection{Residual images}
Point sources in the X-ray images are identified via the \textsc{wavdetect} tool (accounting for the PSF), visually confirmed, and masked.
The exposure-corrected and background-subtracted image is then radially binned into logarithmically-spaced annuli to create a surface brightness profile, $S(r)$, as a function of the projected radius $r$, for each cluster.
We check that X-ray emission from each cluster is observable above the background within the region of interest.
We then perform a least-squares fit of this profile to a spherically symmetric double $\beta$-model:
\begin{equation}
    S(r)=\sum_{n=1}^{2} S_n \left[ 1+\left( \frac{r}{r_{\text{c},n}}\right)^2 \right]^{-3 \beta_n +0.5},\\
    \label{eq:betamod}
\end{equation}
where the free parameters are the surface brightness normalization $S_n$, core radius $r_{\text{c},n}$, and the $\beta_n$ parameter, and $n$ is the number of components (one or two $\beta-$models).
This fitted $\beta$-model is used as the unperturbed {surface brightness model} for each cluster.
For unrelaxed clusters that do not have the excess central emission associated with a cooling flow, the second $\beta$-model component, $S_2$, is fixed to zero. 
To obtain the images of X-ray surface brightness fluctuations (residual images), we divide the initial cluster images by their best-fitting models.

Many clusters exhibit an elliptical X-ray morphology reflecting the underlying gravitational potential of the dark matter halo \citep[e.g.,][]{umetsu_projected_2018,gonzalez_halo_2021}.
We correct our fitted $\beta$-model to better reflect this ellipticity by performing a Levenberg-Marquardt two-dimensional fit of the exposure-corrected image to an elliptical $\beta$-model, where the projected radius $r$ in equation (\ref{eq:betamod}) has been replaced with the elliptical radius $r_\text{e}$, with additional free parameters $\epsilon$ (ellipticity) and $\theta$ (orientation). 
The ratio of the best-fitting elliptical and spherical models is then used to correct the spherically symmetric $\beta$-model described above.

Additionally, the gas density distribution in dynamically disturbed clusters is often asymmetric. 
Assuming a spherically- or even elliptically-symmetric underlying model in such systems can significantly bias the measurement of density fluctuations on large scales.
To model an asymmetric cluster atmosphere, we ``patch'' the $\beta$-model on large spatial scales following \citet{zhuravleva_gas_2015}. 
For details, see Appendix \ref{app:model}.
In short, the residual image is smoothed by a Gaussian with a large width to obtain a correction for the large-scale deviations. 
The unperturbed model (normalized by the image mask smoothed on the same scale) is then multiplied by this correction to get a patched model. 
This ``patched'' model effectively removes the large-scale asymmetries in residual images.
We caution that the choice of patching scale is non-obvious.
In our work, we calculate the power spectrum assuming several patching scales to determine which scale visually removes the asymmetry from the residual image. 
Our analysis only includes measurements of \drrk{} at scales that are unaffected by the choice of patching scale.

\subsection{Power spectral analysis}
\label{sec:powerspec}
To measure the power spectra of density fluctuations from the observed residual images, we follow the analysis steps described by \citet{churazov_x-ray_2012}.
We employ the $\Delta$-variance method \citep{arevalo_mexican_2012} to measure the power spectrum of surface brightness fluctuations, $P_\text{2D}(k)$\footnote{Throughout this work, we adopt the convention that the wavenumber $k$ is the inverse of the spatial scale, without a factor of $2 \pi$. We additionally assume that fluctuations are isotropic, i.e.,  $k=\sqrt{k_x^2+k_y^2+k_z^2}$.}, in the residual image. 
The method is tuned for the non-periodic data with gaps and is best suited for power spectra that are smooth functions. 
{In brief, given a masked image $I$ of the background-subtracted X-ray counts and the exposure map $E$ (product of the \chandra exposure map, unperturbed density model, and mask $M$), the power spectrum ($P_\text{2D}(k)$) of the residual image at a wavenumber $k$ is given by
\begin{equation}
    P_\text{2D}(k) =\frac{\Sigma I_{k}^2}{\Sigma M} \frac{1}{\epsilon^2 \pi k^2},
    \label{eq:variance}
\end{equation}
where
\begin{equation}
    I_{k}=\left( \frac{G_{k_1} \ast I}{G_{k_1} \ast E}-\frac{G_{k_2} \ast I}{G_{k_2} \ast E}\right) M
    \label{eq:mexicanhat}
\end{equation}
with Gaussian kernels $G_{k_1}$ and $G_{k_2}$ of widths $\frac{0.225}{k} \sqrt{1+\epsilon}$ and $\frac{0.225}{k \sqrt{1+\epsilon}}$, where $\epsilon \ll 1$.}

We can deproject the 2D power spectrum of surface brightness fluctuations, $P_\text{2D}(k)$, to a {3D power spectrum of X-ray emissivity fluctuations}, $P_\text{3D}(k)$, and obtain the amplitude of density fluctuations at each wavenumber, i.e., $\left( \delta \rho / \rho \right)_k=\sqrt{4\pi P_\text{3D}(k) k^3}$. 
Assuming that density fluctuations are a homogeneous and isotropic random field, $P_\text{2D}$ is related to $P_\text{3D}$ as
\begin{equation}
    P_\text{2D}(k) \approx 4 P_\text{3D}(k) \int |W(k_\text{z})|^2 dk_\text{z},
    \label{eq:deproj}
\end{equation}
where $W(k_\text{z})$ is the Fourier transform of the normalized emissivity distribution along the line of sight.
The emissivity, and therefore the deprojection factor (integral in equation (\ref{eq:deproj})), depends on the projected radius $r$.
We calculate this deprojection factor at each line of sight and include it in the exposure map from equation (\ref{eq:mexicanhat}).

\defcitealias{zhuravleva_indirect_2023}{Z23}
We measure \drrk{} at each wavenumber $k$ and relate it to the characteristic velocity of subsonic gas motions through the statistical relationship
\begin{equation}
    \left(\frac{\delta \rho}{\rho}\right)_k = \eta \left(\frac{V_\text{1D,k}}{c_\text{s}}\right),
    \label{eq:relation}
\end{equation}
where $V_\text{1D,k}$ is a scale-dependent characteristic one-dimensional velocity of gas motions in the ICM, $c_\text{s}$ is the sound speed and $\eta$ is a proportionality coefficient. 
Recently updated analysis of gas fluctuations in galaxy clusters from cosmological numerical simulations \citep[hereafter Z23]{zhuravleva_indirect_2023} confirmed a strong correlation between velocities and density fluctuations, finding that $\eta$ varies slightly depending on the cluster dynamical state. 
Note that this result differs from the conclusions of \citet{simonte_exploring_2022}, who filter the velocity field in order to separate bulk and turbulent motions.
Given the difficulty in unambiguously defining the filtering scale (both in observations and simulations), we conservatively measure the whole velocity field.
Following \citetalias{zhuravleva_indirect_2023}, {who average $\eta$ between scales of 60 to 300 kpc,} our adopted values for $\eta$ are summarized in Table \ref{tab:propcoeffs}. 
In the near future, this relationship will be tested using direct measurements of the turbulent velocity dispersion with \xrism. 
The first such measurement performed by \textit{Hitomi} was broadly consistent with $\eta \approx 1$ in the Perseus cluster core \citep[]{hitomi_collaboration_atmospheric_2018}.

\begin{table}
    \centering
    \begin{tabular}{c|ccc}
        Dynamical State & Relaxed & Intermediate & Unrelaxed \\
        \hline
        $\eta$ & $0.98\pm0.14$ & $1.04\pm0.18$ & $1.16\pm0.30$ \\
    \end{tabular}
    \caption{Adopted values for the proportionality coefficient $\eta$, from \citepalias{zhuravleva_indirect_2023}. }
    \label{tab:propcoeffs}
    \vspace{-0.5cm}
\end{table}

As our goal is to probe merger-driven gas motions and density fluctuations, we derive the density fluctuation power spectrum within a cluster-centric projected radius of \Rc{2500}, where the effect of the non-X-ray-background is minimized.
We omit (mask) cool-cores where appropriate, as these are often perturbed by AGN activity \citep[e.g.,][]{heinrich_constraining_2021,walker_what_2018} and strong cooling \citep{fabian_cooling_1994,hudson_what_2010}. 
We define $R_\text{cool}$ as the radius within which the gas cooling time $t_\text{cool}$ is less than the age of the universe $t_\text{age}(z)$.

\begin{figure}
    \centering
    \includegraphics[trim=0 15 0 0,width=\linewidth]{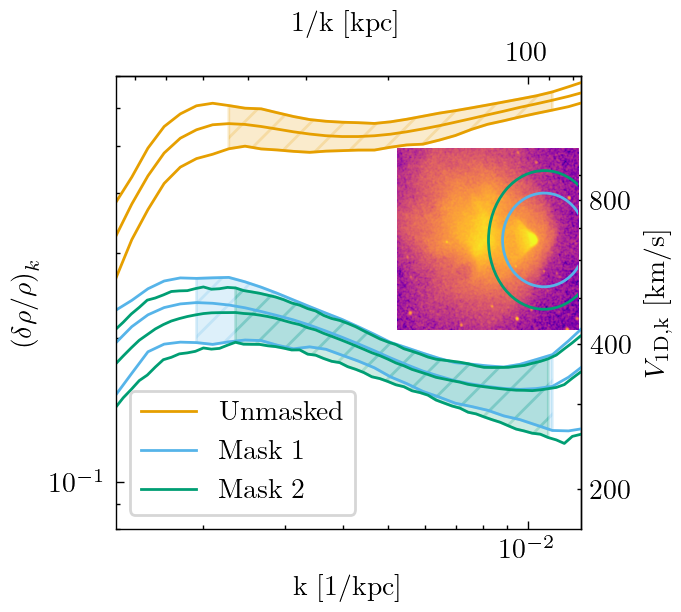}
    \caption{Effect of bright substructures on measurements of \drrk{} in the Bullet cluster. 
    Solid lines show the mean value of \drrk{}, corrected for Poisson noise and effects of the \chandra PSF, with $1\sigma$ uncertainties (upper and lower limits).
    {Shaded regions show wavenumbers where the measurement is unaffected by observational effects.}
    Inset shows the exposure-corrected image of the \Rc{2500} region.
    The orange spectrum is measured in the whole region, while blue and green spectra are measured when their respective ellipses (see inset) are excluded. 
    The exclusion of the blue ellipse is sufficient for characterizing fluctuations in the bulk ICM.}
    \label{fig:masking}
\end{figure}

\subsection{Bright substructures}
\label{sec:substructures}
Galaxy clusters continuously grow through mergers and often exhibit bright substructures that can dominate the signal if not properly excluded.
Our primary goal is to characterize \textit{volume filling (bulk)} gas perturbations. 
As such, for each cluster with prominent substructures, we produce two power spectra: one with substructures included and one without.
Each cluster with identifiable substructures is tested to determine the region size that sufficiently masks each feature.
Figure \ref{fig:masking} provides an example of this test on the Bullet cluster.
This cluster is undergoing a major merger in the plane of the sky that is responsible for the bright cold front (the eponymous ``bullet''), the remnant of the infalling system's cool-core, and the sharp surface brightness discontinuity to the west of the ``bullet'', a bow shock \citep{markevitch_textbook_2002}.
We first calculate \drrk{} for the entire cluster within \Rc{2500}, shown in Figure \ref{fig:masking} as the orange spectrum. 
The blue and green spectra are then calculated after masking their respective regions shown on the inset.
One can see the spectrum significantly decreases when the blue region is masked, but shows little change in the case of the larger green region. 
Therefore, we conclude the blue region is sufficient to remove the effect of these substructures on the spectrum as a whole.
The slight difference between the blue and green spectra at large spatial scales is due to the lower sampling of these scales when the region size is increased.
In our final analysis, we only include scales that are largely unaffected by the choice of masked region.

\subsection{Limits on observable scales}
\label{sec:limits}
Each cluster must be carefully examined to determine which length scales are reliably probed within \Rc{2500}, given the available data.
On large scales, our measurement is limited by the region size.
When $1/k$ approaches this radius, the number of large-scale modes decreases significantly, causing the spectrum to flatten and decrease around $k<1/R_{2500}$.
This problem of ``sample variance'' is treated in detail by \citet{dupourque_investigating_2023}.
In this work, we conservatively limit our measurements to scales below this peak, and caution that this minimum wavenumber does not necessarily represent the turbulent injection scale.

On small scales, the observations are limited by Poisson noise, unresolved point sources, and suppression of structures due to the \chandra PSF. 
Note that though the PSF is nominally quite small, Gaussian smoothing in the $\Delta-$variance method makes the PSF effect noticeable at larger scales than the nominal PSF size.

The Poisson uncertainty associated with the number of X-ray counts per pixel contributes a scale-independent component to the measured $P_\text{3D}$ of order $\Sigma I_\text{raw}/\Sigma E^2$, summing over the raw X-ray counts image and the square of the exposure map.
We can accurately model this contribution by creating many (in practice, $N_\text{pois}=60$) realizations of random images scaled by the Poisson uncertainty, the square root of the X-ray counts/pixel. 
The average power spectrum of these randomized images is subtracted from the nominally measured $P_\text{2D}$ or $P_\text{3D}$ and used to inform the $1\sigma$ uncertainties.

While we detect and remove visible point sources from our data, there is likely a contribution of unresolved point sources to the power spectrum on small scales.
Following \citet{churazov_x-ray_2012}, we attempt to quantify the power spectral contribution of these unresolved point sources.
We first measure the power spectra of bright, identifiable point sources, $P_\text{bright}$, by calculating the difference between power spectra when they are excised and when they are included.
This $P_\text{bright}$ can then be rescaled to estimate the contribution of undetected point sources based on their flux distribution.
In each cluster, we measure the background-subtracted flux of each point source, determining the number of sources per flux bin $dN/dF$.
Between the brightest point source in the field ($F_\text{max}$) and the dimmest ($F_\text{min}$), this distribution follows a power law with respect to F. 
It is then reasonable to conclude that $P_\text{bright}$ is proportional to the total flux of these point sources, namely,
\begin{equation}
    P_\text{bright} \propto \int\limits_{F_\text{min}}^{F_\text{max}} \frac{dN}{dF} F^2 dF.
\end{equation}
By definition, unresolved point sources occupy fluxes between $F=0$ and $F=F_\text{min}$.
Assuming that the slope of the flux distribution does not change in this region, we can find the relative contribution of unresolved point sources by numerically integrating
\begin{equation}
   P_\text{faint} \propto \int\limits_{0}^{F_\text{min}} \frac{dN}{dF} F^2 dF.
\end{equation}
The ``flux ratio'' between these two integrals is calculated for each cluster and multiplied by $P_\text{bright}$ to find $P_\text{faint}$. 
We then limit our measurements to scales where $P_\text{3D}\gg P_\text{faint}$.

Finally, we consider the smoothing of small-scale fluctuations by the \chandra PSF.
As the $\Delta$-variance method calculates a low-resolution, smoothed power spectrum, it is not sufficient to limit our measurements to scales greater than $\sim$ a few arcseconds.
Instead, we create mock observations of many randomly placed point sources in the region of interest and measure two power spectra: one of the raw image where each point source occupies one pixel, and one of the point sources after they have been convolved with the exposure-weighted \chandra PSF.
The ratio of these, $R_\text{PSF}=P_\text{conv}/P_\text{raw}$, represents the fractional effect of the point spread function on our measured $P_\text{3D}$.
We calculate this correction several ({10}) times, each time generating a new map of randomly-distributed point sources, and take the average values of $R_\text{PSF}$ at each wavenumber.
We divide the Poisson-subtracted $P_\text{3D}$ by this correction to obtain the suppressed, high-$k$ power, and limit our measurement to scales where $1-R_\text{PSF}<25$ per cent.

After all the corrections and checks of systematic uncertainties, we find a range of scales for each cluster where our power spectra measurements are most reliable.
These scales are listed in Table \ref{tab:clumpings} and shown in Figure \ref{fig:allspectra}.
The median cluster in our final sample is observable between $\sim 120$ and $\sim 360$ kpc, spanning a range of $k_\text{max}/k_\text{min}\approx 3$.
Most clusters ($N=68$) are observable between scales of 200 to 300 kpc. 
In our sample, 24 clusters are measurable at scales less than 100 kpc.
The smallest scales in the sample ($1/k \approx 60$ kpc) are achieved in A3667 and A1367.

\section{Results}
\label{sec:results}

\subsection{Density fluctuations and clumping factors}
\label{sec:fluctuations}

\begin{figure*}
    \centering
    \includegraphics[width=\textwidth]{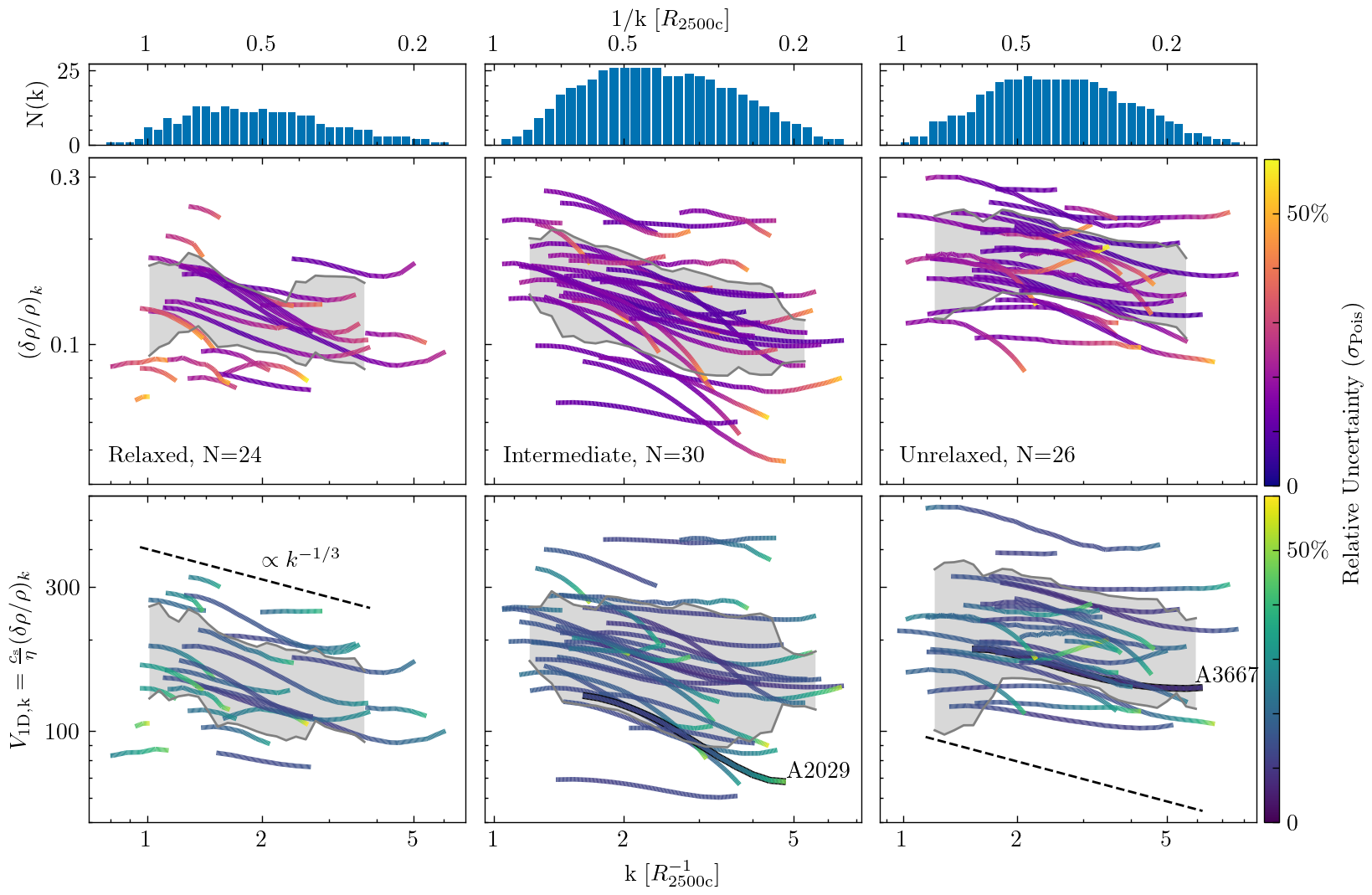}
    \vspace{-0.5cm}
    \caption{\textbf{Middle row:} Amplitude of gas density fluctuations vs. wavenumber (bottom x-axis) and length scale (top x-axis), in units of \Rc{2500}. Each line represents one cluster, where \drrk{} is measured within a region of radius \Rc{2500} (excluding cool-cores). Clusters are divided into relaxed (left), intermediate (middle), and unrelaxed (right) subsamples. Only scales largely unaffected by systematics (as discussed in Section \ref{sec:limits}) are shown. Line color corresponds to the measurement's relative uncertainty, derived from $1 \sigma$ Poisson uncertainties (see color scale at right). Subsample-averaged amplitudes at each length scale, plus/minus one standard deviation, are shown in grey. \textbf{Bottom row:} Velocity power spectra for each cluster, with equivalent notations. The Kolmogorov slope is plotted as a guide (dashed black line). Two highlighted clusters, A2029 and A3667, will be observed with \textit{XRISM} during the PV phase. \textbf{Top row:} Number of clusters with measured \drrk{} at each scale within each subsample. Prominent substructures are removed from the analyzed images to measure fluctuations in the bulk ICM. For the version with included substructures, see Figure \ref{fig:allspectra_str}.}
    \label{fig:allspectra}
\end{figure*}

We measure the scale-dependent amplitude of density fluctuations \drrk{} and velocity power spectrum $V_\text{1D,k}$ in a total of 80 galaxy clusters across length scales ranging from $50$ kpc $\lesssim 1/k \lesssim 750$ kpc.
Measurements from the bulk ICM (substructures excluded) are shown in Figure \ref{fig:allspectra}, where the sample is divided into relaxed, intermediate, and unrelaxed clusters.
In this figure, we plot \drrk{} and $V_\text{1D,k}$ against wavenumber/spatial scale in units of \Rc{2500} (bottom/top axis), where line color corresponds to the relative uncertainty $\sigma_{\left(\delta \rho / \rho \right)_k}/$\drrk{} with $1\sigma$ uncertainties derived from Poisson noise as described in Section \ref{sec:limits}. 
Weighted sample averages for each morphological category are additionally shown in grey, with the shaded region corresponding to the average $\pm 1$ standard deviation.
The top panels show the number of clusters with measured \drrk{} at a given scale.
Figure \ref{fig:allspectra_str} shows equivalent measurements taken from the entire cluster, with substructures included in the calculation. 
We stress that the latter measurements do not represent fluctuations in the bulk ICM.

Typical amplitudes of density fluctuations in the bulk ICM are $\sim 12 \pm 2$ per cent for relaxed clusters, increasing to $\sim 19 \pm 5$ per cent for unrelaxed clusters when measured at $1/k=0.5$ \Rc{2500} (Figure \ref{fig:allspectra}). 
If substructures are included in the measurement, the level of fluctuations in relaxed clusters remains about the same, while the typical amplitude increases to $\sim 40$ per cent in unrelaxed clusters \ref{fig:allspectra_str}). 
One would expect to see increasing density fluctuations with merger activity, and while this is the case {in the entire region} (Figure \ref{fig:allspectra_str}), the effect is smaller once prominent substructures are removed.

{The difference in amplitudes in the bulk ICM translates to a factor of $\gtrsim 2.5$ more ICM kinetic energy at length scales smaller than $0.5$\Rc{2500} in unrelaxed clusters compared to relaxed ones.
This suggests that fluctuations in the bulk ICM are not strongly linked to the current merger state of the cluster.
This may be because there has not been sufficient time for fluctuations in clusters with ongoing mergers to cascade to these relatively small scales.
Another possibility is that fluctuations in the bulk ICM of unrelaxed clusters dissipate quickly.
The opposite may also be true, that those fluctuations present in relaxed clusters are ``left over'' from mergers in the cluster's more distant history.}

It is additionally interesting to assess whether the density fluctuation amplitudes in relaxed vs. unrelaxed clusters belong to different populations.
A Kolmogorov-Smirnov test performed at each wavenumber shows that it is difficult to distinguish between relaxed and intermediate clusters. However, density fluctuations measured in unrelaxed clusters (including substructures) are distinct from relaxed and intermediate clusters at the $p<0.05$ level between scales of $1/k \approx 0.2$ to $0.8$ \Rc{2500}. 
Therefore, at these wavenumbers, the amplitude of density fluctuations could be used as an indicator of cluster dynamic state, complementing the commonly used indicators based on, e.g., the asymmetry of the X-ray surface brightness, centroid shifts, power ratios, and symmetry-peakiness alignment \citep[e.g.,][]{cerini_new_2022, yuan_dynamical_2022, cui_dynamical_2017, mantz2015}. 
{Although there is some overlap between the intermediate and unrelaxed populations, \drrk{}$>20$ per cent at $1/k=0.5$\Rc{2500} is typically indicative of a major merger.}

\begin{figure}
    \centering
    \includegraphics[width=\columnwidth]{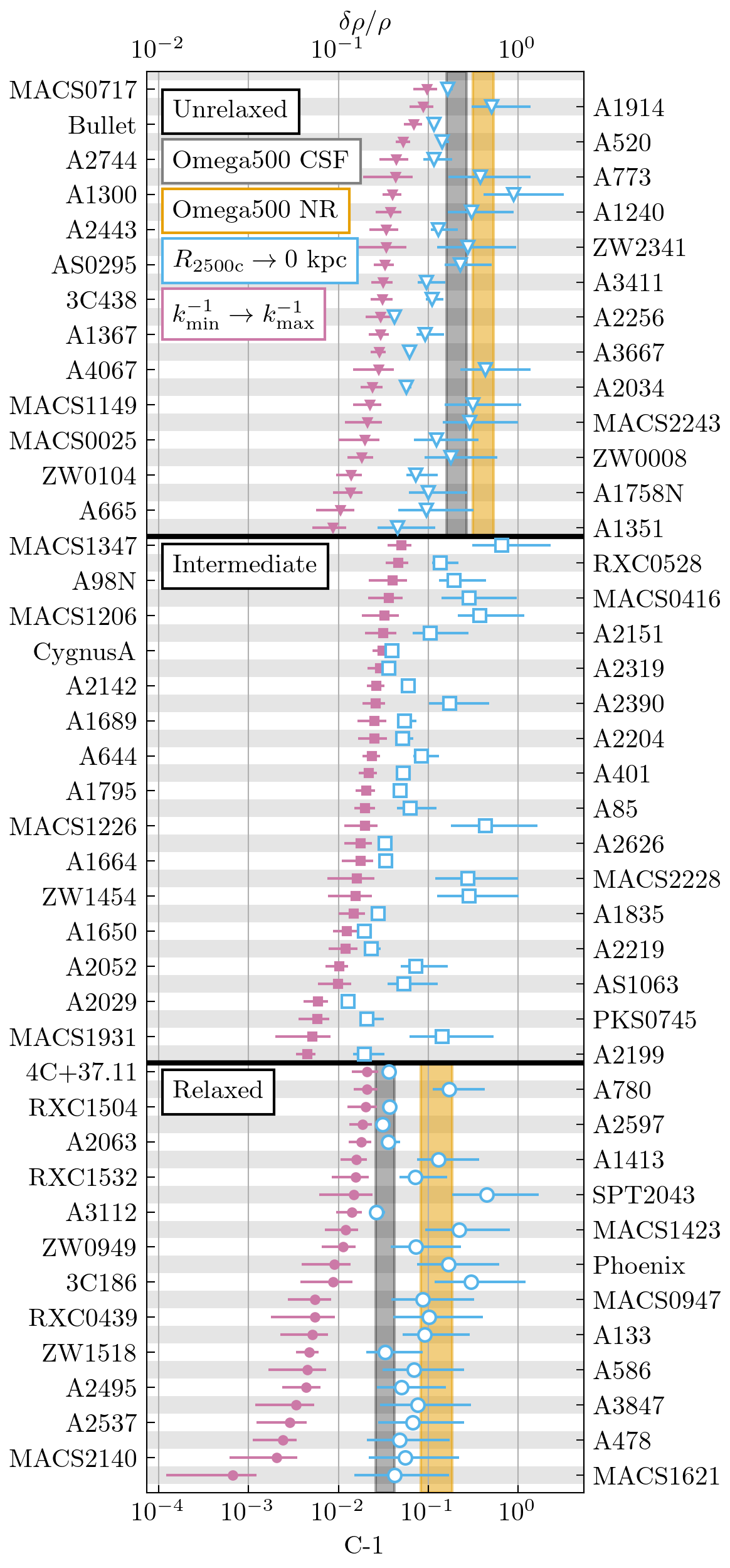}
    \vspace{-0.7cm}
    \caption{ICM clumping factors obtained by integrating \drrk{} for each cluster in our sample. In pink, we integrate equation (\ref{eq:clumping}) from $k_\text{min}$ to $k_\text{max}$, providing a lower limit on each value.
    We additionally fit the observed spectrum to a power-law via MCMC and integrate from $k=1/$\Rc{2500} to $k=\infty$.
    The median values, along with upper and lower uncertainties, are presented in blue with unfilled markers.
    These values could be interpreted as upper limits.
    Clumping factors measured in the Omega500 cosmological simulations \citep{nagai_gas_2011,zhuravleva_quantifying_2013} are plotted in shaded regions in their corresponding morphological categories. Grey regions show simulations where radiative processes (e.g., cooling and star formation/feedback) are included, while yellow regions show results from non-radiative simulations.}
    \label{fig:clumpings}
\end{figure}

The scale-dependent amplitude of density fluctuations \drrk{} can additionally be used to provide constraints on the gas clumping factor, $C$, in the ICM through a simple relation \citep{zhuravleva_gas_2015}:
\begin{equation}
    \label{eq:clumping}
    C-1 =\left( \delta \rho / \rho \right) ^2=\int \left( \frac{\delta \rho }{\rho} \right)_k^2 \frac{dk}{k}. 
\end{equation}
This expression should be integrated over the entire range of wavenumbers, however, our measurements are limited to a smaller range of scales described in Section \ref{sec:limits} and listed in Table \ref{tab:clumpings}.
Therefore, we integrate across this range of measurable scales, interpreting this result as a lower limit on $C$, which we plot in pink in Figure \ref{fig:clumpings}.

To estimate the upper limit on $C$, we fit \drrk{} to a power-law via an MCMC chain \citep{hogg_data_2018} using the \textsc{emcee} package \citep{foreman-mackey_emcee_2013}, and integrate this power-law from $k=1/$\Rc{2500} to $k=\infty$. 
{An example of a fitted power spectrum is shown in Figure \ref{fig:mcmcfit}.}
The choice of $k=1/$\Rc{2500} is rather arbitrary, however, reasonable given the size of the considered region. 
These upper limits are plotted as unfilled blue markers in Figure \ref{fig:clumpings}, with uncertainties derived from the 84th and 16th percentile of the MCMC chain.
Both lower and upper limits on the clumping factor, as well as the best-fit parameters of this power-law, i.e., the normalization of \drrk{} at $1/k=$\Rc{2500} and the slope $\alpha$, are presented in Table \ref{tab:clumpings}.

The weighted-average median upper limit on the clumping factor in our sample is $C \approx 1.05 \pm 0.04$.
In relaxed and intermediate clusters, the average is $1.04 \pm 0.02$, compared to $1.09 \pm 0.04$ in unrelaxed clusters.
We find reasonably good constraints on $C$ in 23 clusters, where the median lower limit on $C-1$ is within a factor of 3 of the upper limit.
Namely, these clusters are Cygnus A, A1795, A2597, the Bullet cluster, A2219, A3667, A2319, A2142, A2034, A1664, A1650, MACS0717, A401, A1835, A1689, A2256, RXC1504, A3112, A2626, A2204, 4C+37.11, A2029, and A2063.
Of these, most of the median upper limits are within the $1.02 \lesssim C_\text{U} \lesssim 1.06$ range. On the high end are the Bullet cluster ($C_\text{U} \approx 1.12$) and MACS0717 ($C_\text{U} \approx 1.17$), objects with (likely) the highest clumping factors in our sample.
A1914 and RBS1316 also have high $C_\text{L}$ but have significantly larger uncertainties.
We find the lowest median upper limit on the clumping factor ($C_\text{U} \lesssim 1.02$) in A1650, A2199, and A2029, despite the presence of sloshing in A1650 and A2029. 
{In comparison with other observations, our measurements are consistent with those of \citet{eckert_gas_2015}, who found $C\approx 1.05 \pm 0.04$ within $0.5$\Rc{500} in 31 clusters. 
We only confirm a single cluster with $C$ outside of this range (MACS0717).}

\subsection{Velocity power spectra and non-thermal pressure}
\label{sec:velocities}
We convert our measured \drrk{} into velocity power spectra using equation (\ref{eq:relation}).
{We estimate the median sound speed in the region of interest from either our measured temperature profiles or those found in the literature, shown in Table \ref{tab:parameters}.}
The proportionality coefficient $\eta$ has been recently updated by \citetalias{zhuravleva_indirect_2023} (see their Figure 7) using a sample of 78 clusters from non-radiative cosmological simulations.
In that work, the amplitude of density fluctuations and the Mach number of gas motions were calculated with an equivalent methodology to the one used here within a similar region.
They find average values of $\eta=(\delta\rho/\rho)_k/\mathcal{M}_{\rm 1D,k}$ between $1/k=60$ and $300$ kpc, where the simulation resolution is not expected to affect the measurement. 
The $\eta$ values we use are summarized in Table \ref{tab:propcoeffs}.
Resulting velocity power spectra ($V_\text{1D,k}$) are presented in Fig \ref{fig:allspectra}. 
Additionally, velocities in each morphological category are averaged in the same fashion as $\delta \rho / \rho$, plotted in grey regions with uncertainties as in the middle row. 
We see that, on average, $V_\text{1D,k}$ is correlated to cluster dynamical state. 
On a length scale of $0.5$ \Rc{2500}, the typical $V_\text{1D}$ is $\sim 150$ km s$^{-1}$ for relaxed clusters, $\sim 200$ km s$^{-1}$ for intermediate clusters, and $\sim 250$ km s$^{-1}$ for unrelaxed clusters. 
The velocity amplitude decreases with scale, and the slope is consistent with the Kolmogorov model within the uncertainties.

To estimate the characteristic (i.e., scale-independent) velocity associated with these gas motions, we use the integrated \drr{} and calculate the one-dimensional Mach number, $\mathcal{M}_\text{1D}= (\delta \rho / \rho )/ \eta$.
This measurement can additionally be used to calculate the kinetic pressure fraction:
\begin{equation}
    \label{eq:pressure}
    P_\text{kin}/(P_\text{kin}+P_\text{thm})=P_\text{kin}/P_\text{tot}=\frac{V_\text{1D}^2}{V_\text{1D}^2+kT/\mu m_p}=\left( 1+\frac{1}{\gamma \mathcal{M}_\text{1D}^2}\right)^{-1}
\end{equation}
where $P_\text{kin}$ is the kinetic gas pressure, $P_\text{thm}$ is the thermal pressure, $P_\text{tot}$ is the total pressure, and $V_\text{1D}=c_s \mathcal{M}_\text{1D}$ is the characteristic one-dimensional velocity, assuming an adiabatic index of $\gamma=5/3$. 
{As stated previously, we do not attempt to distinguish between bulk ($V_\text{bulk}$) and turbulent ($V_\text{turb}$) gas motions.
To calculate $P_\text{kin}/P_\text{tot}$, we assume that $V_\text{1D} \sqrt{3} = V_\text{3D,eff}= \sqrt{3 V_\text{turb}^2 + V_\text{bulk}^2}$.}
High-resolution X-ray spectroscopy from \xrism, \textit{NewAthena}, and \textit{LEM} \citep{xrism_science_team_science_2020, barret_athena_2020, kraft_line_2022} will be able to distinguish these components to confirm our measurements and this relationship.

\begin{figure*}
    \centering
    \includegraphics[width=\textwidth]{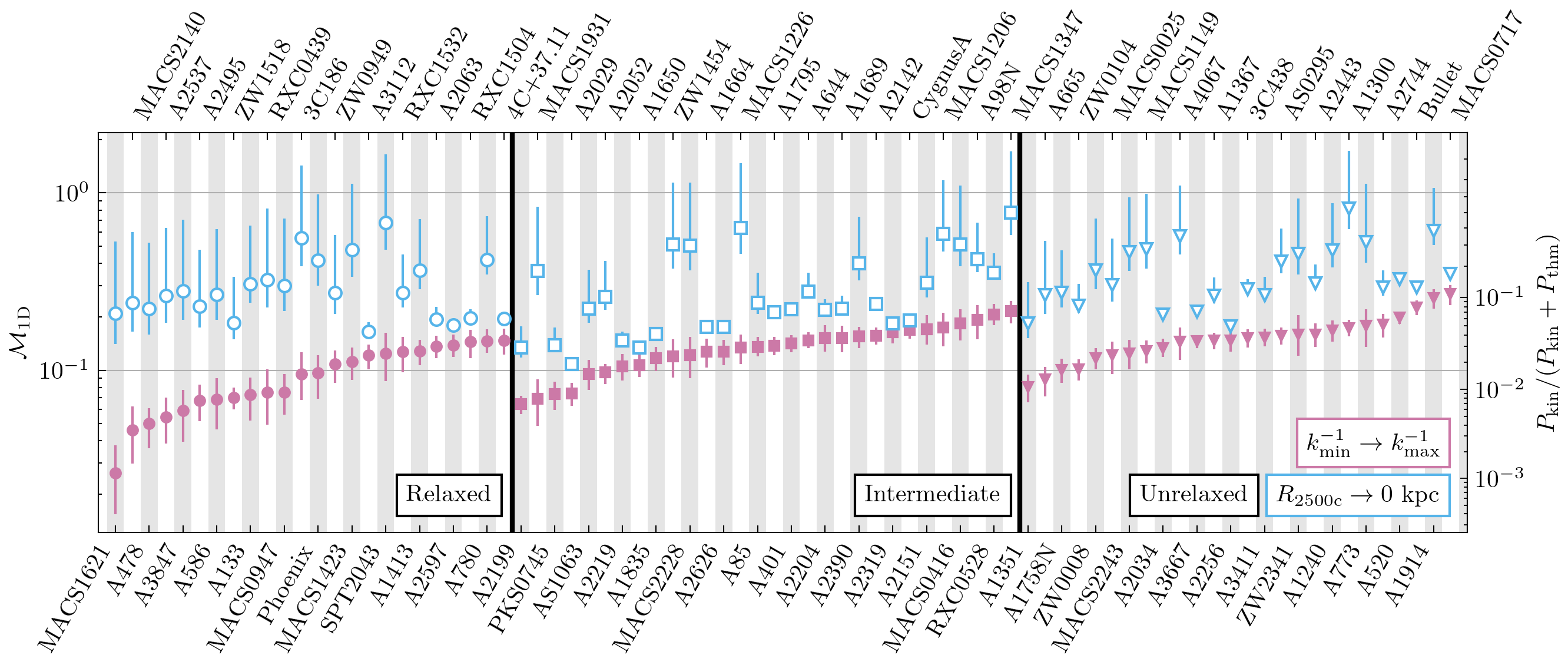}
    \caption{Characteristic one-dimensional Mach numbers (left y-axis) and kinetic pressure fractions (right y-axis) for each cluster in our sample. Notations are equivalent to Figure \ref{fig:clumpings}.}
    \label{fig:machnumbers}
\end{figure*}

Figure \ref{fig:machnumbers} shows the derived values for $\mathcal{M}_\text{1D}$ and $P_\text{kin}/P_\text{tot}$ for all clusters in our sample (see also Table \ref{tab:clumpings}) with equivalent notation to Figure \ref{fig:clumpings}.
We find an average best-fit Mach number in the range of $0.18 \pm 0.05$ across the entire sample, consistent with subsonic gas motions.
The highest characteristic velocities we can confirm are found in MACS0717 ($V_\text{1D}\gtrsim 575$ km s$^{-1}$, $\mathcal{M}_\text{1D} \gtrsim 0.27$) and RBS1316 ($V_\text{1D}\gtrsim 435$ km s$^{-1}$, $\mathcal{M}_\text{1D} \gtrsim 0.22$). 
In all respects (lower limits and best-fit clumping factor/characteristic velocity), this makes RBS1316 {\citep[a.k.a RX J1347.5-1145][]{johnson_sloshing_2012}} the most disturbed cool-core cluster, indicating that significant merger effects are taking place.

The lowest velocities are found in A2597, A1650, A2626, A2199, and A2029, with upper limits on $V_\text{1D}<200$ km s$^{-1}$, all of which are cool-core clusters.
Of these, A1650, A2626, and A2029 show some evidence of mergers in the form of sloshing, while the rest are significantly relaxed.
For non-cool-core clusters, the lowest velocities we confirm are A3667 and A2256, with upper limits below $V_\text{1D}=300$ km s$^{-1}$. 
Both clusters have significant substructures in the form of cold fronts \citep[e.g.,][]{de_gasperin_meerkat_2022,rajpurohit_deep_2023} and bow shocks in the case of A3667.
Confirmation of the characteristic velocities in the bulk ICM with direct observation should carefully exclude emission from these prominent substructures.

Measuring non-thermal pressure support is important for cosmology based on hydrostatic cluster mass measurements \citep[e.g.,][]{allen_improved_2008, vikhlinin_chandra_2009, mantz_cosmology_2016,clerc_x-ray_2023}.
We place upper limits (denoted with the lower index $U$) on the fraction of kinetic pressure in the bulk ICM at $ \left( P_\text{kin}/P_\text{tot}\right)_U<10$ per cent in a total of 22 clusters within the \Rc{2500} region, with $\left(P_\text{kin}/P_\text{tot}\right)_U< 5$ per cent in A2219, A1650, PKS0745, A1835, A2199, and A2029.
A2029 has the lowest non-thermal pressure contribution  of $\sim 1.9 $ per cent.
Most clusters among these 22 are classified as relaxed or intermediate.
Three clusters in this subsample, however, are unrelaxed clusters: A3667, A2034, and A2256, each with $\left( P_\text{kin}/P_\text{tot} \right)_U< 8 $ per cent in the bulk ICM despite significant ongoing  merger processes.
For these clusters, it is important to note that the measured kinetic pressure in the bulk ICM is unlikely to reflect the total kinetic pressure in the entire cluster, contributions from merger-driven substructures could be significant.
In our sample, the weighted-average best-fit upper limit for the kinetic pressure fraction ($\langle P_\text{kin}/P_\text{tot} \rangle_U$) are approximately $6 \pm 2$ per cent for relaxed clusters, $4 \pm 2 $ per cent for intermediate clusters, and $7 \pm 4 $ per cent for unrelaxed clusters.
Note that intermediate and unrelaxed clusters will likely have significant kinetic pressure associated with discrete substructures, which are removed from these measurements. 

{Recent constraints have been placed on the kinetic pressure fraction in a similar region of the Perseus \citep{de_vries_chandra_2023} and Ophiuchus \citep{gatuzz_measuring_2023} clusters, finding $6\lesssim P_\text{kin}/P_\text{tot}\lesssim 13$ per cent and $P_\text{kin}/P_\text{tot} \lesssim 18$ per cent, respectively.
All of our clusters agree with the latter measurement, and all but 12, with upper limits on $P_\text{kin}/P_\text{tot} < 6$ per cent, agree with the former.}

\section{Discussion}
\subsection{Comparisons with cosmological simulations}
\subsubsection{Multi-scale density fluctuations}
\label{sec:simulations}
 
\begin{figure*}
    \includegraphics[width=\textwidth]{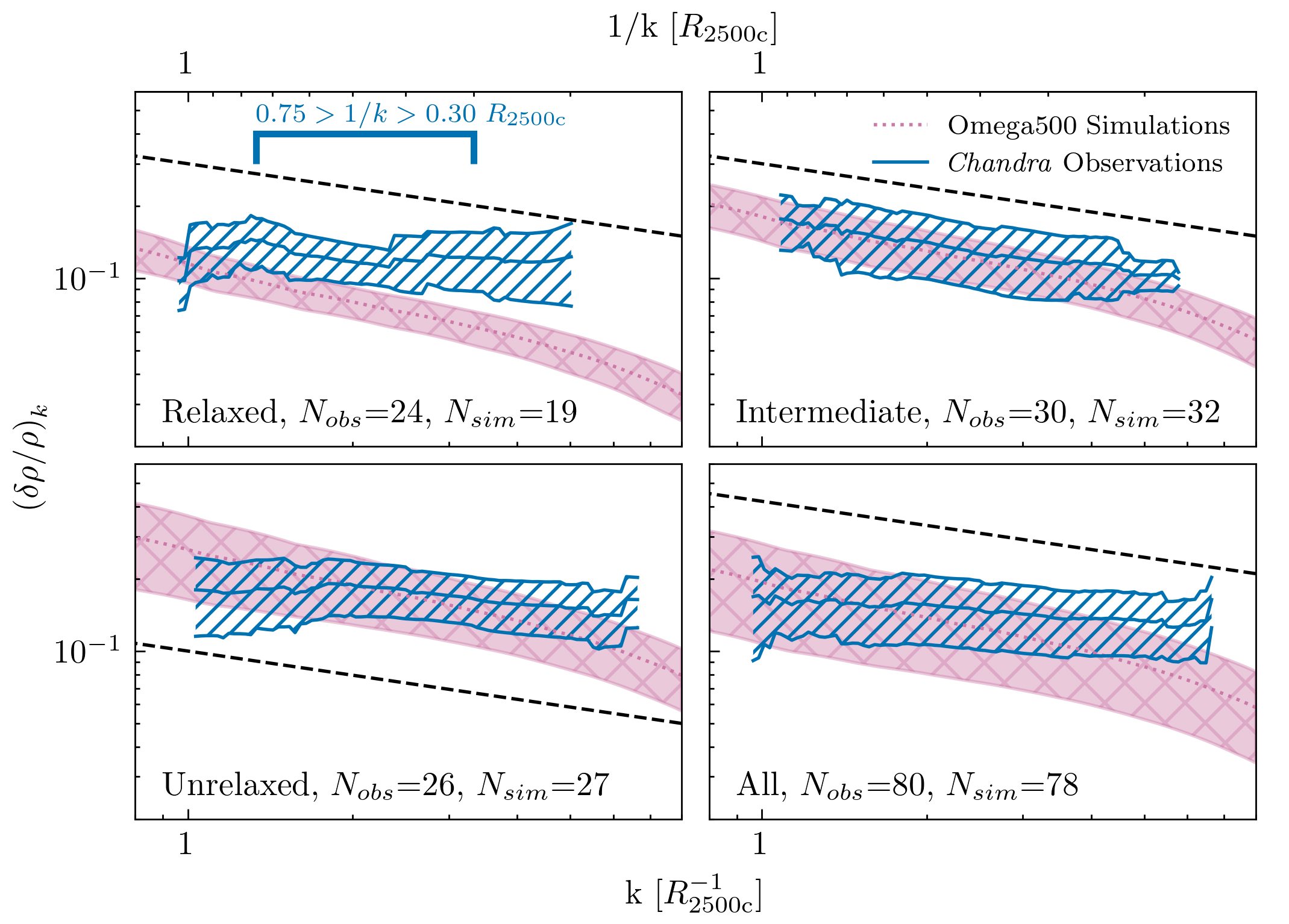}
    \vspace{-0.6cm}
    \caption{Sample-averaged \drrk{} measured in the bulk ICM of relaxed, intermediate, unrelaxed, and all clusters. Observational measurements from this work are plotted in blue, with $1\sigma$ uncertainties in the hatched regions. Measurements from \textsc{Omega500} cosmological simulations \citep[for power spectra, see][]{zhuravleva_indirect_2023} are plotted in pink. The blue bracket represents length scales where we confirm a divergence between observed and simulated measurements in relaxed systems. Black dashed lines represented a $k^{1/3}$ power-law relationship for guidance.}
    \label{fig:simcomparison}
\end{figure*}

Scale-dependent amplitudes of density (or any other characteristic) fluctuations can be used as sensitive probes of multi-scale, non-linear physics and sub-grid prescriptions in numerical simulations. 
An equivalent method of measuring density fluctuations has been recently applied to a representative sample of 78 galaxy clusters \citep{zhuravleva_indirect_2023} from the \textsc{Omega500} cosmological simulations \citep{nagai_testing_2007, nagai_effects_2007,nelson_hydrodynamic_2014}. 
Similar to our work, the authors split the simulated clusters into relaxed, intermediate, and unrelaxed systems and considered fluctuations in the bulk ICM (i.e., removing prominent substructures) within a similar region ($r \leq 0.5 R_{500c}$).
The authors considered fluctuations in simulations with non-radiative physics, which apply to the regions considered in our work (i.e., beyond cool-cores and least affected by cooling and feedback processes). 

Figure \ref{fig:simcomparison} shows the average amplitude of density fluctuations, with one standard deviation as uncertainties, in simulated and observed clusters in pink and blue, respectively. 
In {clusters with evidence of mergers} (the intermediate and unrelaxed subsamples), observed and simulated amplitudes of density fluctuations are consistent across all scales.
In relaxed clusters, however, the amplitudes diverge on small scales, reaching a factor $\gtrsim 2$ difference between the mean values at $1/k \lesssim 0.3$\Rc{2500} {(typically $\sim $ 120 kpc)}.
Performing a Kolmogorov-Smirnov test at each length scale, we confirm that the distribution of density fluctuations observed in our relaxed cluster sample are distinct from the simulations between $1/k =0.75$ and $0.3$ \Rc{2500} ({$\sim 300 - 120$ kpc,} the dark blue bracket in the top-left panel), with the probability of the two samples being derived from the same distribution being $<0.05$ at these scales. 
On scales smaller than $1/k \sim 0.3$\Rc{2500}, we observe a similar tension, however, the number of clusters where we could probe these scales is small, less than 10. 
{If the observational and simulated measurements are compared with wavenumber in kpc$^{-1}$ (instead of re-scaling with \Rc{2500}), we observe this tension in relaxed clusters between $\sim$120 and 300 kpc. }

We checked whether observational biases could cause this tension. 
Fluctuations in five clusters in the relaxed sample (A2537, A3847, A478, MACS1621 and MACS2140, see Figure \ref{fig:allspectra}) are measured on large scales ($1/k \gtrsim 0.6$\Rc{2500}) with \drrk{}$<10 $ per cent. 
Extrapolating these amplitudes to smaller scales gives very small amplitudes of fluctuations, which are difficult to measure reliably given Poisson noise, background, and contribution from unresolved point sources. 
However, if these spectra were observed on small scales, they could reduce the average amplitude of observed density fluctuations, bringing it closer to the simulation average. 
To examine this potential bias, we extrapolate the power spectra in these five clusters across smaller scales, fitting the observed spectra with a power-law as described in Section \ref{sec:fluctuations}, and assuming the slope of \drrk{} extends across the entire range of length scales. 
With these extended power spectra, we reanalyze the tension in relaxed clusters. 
While the distributions of \drrk{} in the observed and simulated subsamples become compatible at small scales (where we initially observed less than 10 clusters), the tension remains strong between  $0.3 \lesssim 1/k \lesssim 0.75$ \Rc{2500}. 
This conclusion remains if we assume a fixed Kolmogorov slope of $-1/3$, or if the extrapolation across all scales is applied to every cluster in the sample.

Our measured power spectra could also be affected by unresolved point sources or, e.g., additional variations on the detector, especially on small scales. 
To check the potential effect of these, we take a more conservative value for the maximum wavenumber measured reliably in each cluster {by only considering wavenumbers where we do not observe any flattening of the \drrk{} slope}.
We find that even with reduced limits, the divergence between observation and simulations remains across most of the range.
{We additionally compare the observed spectrum when no correction for the \chandra PSF (see Section \ref{sec:limits}) is applied, and find no change in the length scales where the divergence is observed.}

One could expect that density fluctuations also correlate positively with redshift since the merger rate increases with redshift according to hierarchical structure formation. 
Our sample covers a range of redshifts up to $z \sim 1$, while the \textsc{Omega500} clusters were analyzed at $z=0$. 
To check whether the observed tension could be caused by this redshift evolution of $\delta \rho / \rho$, we recalculate the Kolmogorov-Smirnov test excluding high-redshift systems, and conclude that the redshift evolution did not significantly affect the result.

Finally, we check if our morphological classification of galaxy clusters has any effect on the observed tension between the observations and simulations. 
We identify clusters that may be on the boundary between morphological categories and recalculate the averaged spectra and K-S test when the morphological classification is randomly varied between categories for each boundary object.
We find that the tension is strong even if the classification is varied or if these boundary objects are removed from the sample entirely. 
To conclude, the tension appears real and is stable to observational systematics. 
{Simply put, we do not observe clusters with the low-amplitudes fluctuations that are common in simulations at scales of $1/k \lesssim 0.75$\Rc{2500}}.

Note, the displayed scales from simulations are not expected to be affected by the resolution of simulations, which becomes evident on length scales below $\sim 60$ kpc \citepalias[$\lesssim 0.15$\Rc{2500},][see their Section 4.2]{zhuravleva_indirect_2023}.
Moreover, the divergence in averaged \drrk{} is only observed in relaxed clusters and not in the intermediate or unrelaxed subsamples.
{Therefore, this divergence in relaxed clusters is unlikely to be driven {primarily} by numerical viscosity.}

As the factor of $\sim 1.5$ to $2.5$ difference between the amplitude of density fluctuations on length scales smaller than $0.75$\Rc{2500} in relaxed clusters cannot be explained by known observational/instrumental effects and is unlikely to be caused by the resolution of cosmological simulations, it could be attributed to missing physics in the simulations we consider.
Radiative processes (e.g., gas cooling), AGN feedback, galactic outflows, and magnetic fields could all increase \drrk{} at these scales. 
Fluctuations in relaxed clusters appear to be more sensitive to these physics compared to perturbed systems, in which ongoing mergers continuously ``stir'' the ICM and contribute significantly to the observed amplitude of density fluctuations. 
While it is straightforward to extend the analysis of density fluctuations to other simulations and simulation runs that include these additional physics, it is beyond the scope of this work.
The observed tension also means that relaxed clusters should be selected for testing additional physics implemented in simulations, especially multi-scale physics, against observations.

{With these measurements, we cannot distinguish between which of the above processes is the primary driver of the observed divergence.
The central AGN {mostly affects} gas dynamics within the cool-core and is not influential in the regions considered, however AGN and galaxy-scale outflows originating from other cluster galaxies could affect density fluctuations in relaxed clusters more strongly than in merging ones.
Additionally, while cooling times in the regions considered are greater than the Hubble time, radiative physics could still affect \drrk{} in simulations {(e.g., by creating clumpy structures, see Section \ref{sec:simclumping} below).}}

{Finally, the fact that relaxed clusters appear ``smoother'' in simulations compared to observations could have consequences for cluster cosmology predictions based on hydrostatic mass measurements  \citep[e.g.,][]{mantz_cosmology_2015,vikhlinin_chandra_2009,pratt_galaxy_2019}. 
Namely, the predicted contribution of non-thermal pressure from simulations could be underestimated \citep[e.g.,][]{lau_constraining_2012, nelson_hydrodynamic_2014}. 
We leave the quantitative exploration of this issue for future works.}

\subsubsection{Clumping factor and non-thermal pressure}
\label{sec:simclumping}
It is interesting to compare the estimated ICM clumping factor shown in Figure \ref{fig:clumpings} with predictions from numerical simulations. 
In earlier works based on \textsc{Omega500} simulations, \citet{nagai_gas_2011} found that $1.03<C<1.04$ in relaxed clusters and $1.16<C<1.26$ in unrelaxed ones within \Rc{2500} when radiative processes (e.g., cooling, star formation/feedback, hereafter CSF) are included  \citep[see also][]{zhuravleva_quantifying_2013}. 
The non-radiative runs predict a higher level of clumpiness, $\sim 1.14$ ($1.48$) for relaxed (unrelaxed) systems, see gray and yellow regions in Figure \ref{fig:clumpings}. 
More recently, \citet{angelinelli_proprieties_2021} found $1.05<C<1.15$ in a sample of clusters from the non-radiative Itasca cluster sample \citep{vazza_turbulence_2017}. 
Our measurements are broadly consistent with these values. 
Specifically, we find that nearly all relaxed and two-thirds of unrelaxed clusters in our sample are consistent with the \textsc{Omega500} CSF predictions. 
The remaining third of unrelaxed clusters have upper limits lower than predicted in these simulations. 
Clumping in non-radiative runs in \textsc{Omega500} is systematically higher and consistent with $\sim 70$ per cent of the relaxed and $\sim 45$ per cent of unrelaxed clusters from our sample. 
As for the Itasca simulations, we find that 15 out of 80 clusters in our sample have clumping factors constrained below $1.05$, the rest are consistent with the predicted range.

The non-thermal pressure fraction, shown in Figure \ref{fig:machnumbers}, has also been measured in several cosmological simulations.
\citet{battaglia_cluster_2012,nelson_hydrodynamic_2014,shi_analytical_2015} find that, broadly, $P_\text{kin}/P_\text{tot}$ ranges between $\sim 5 $ per cent and $25 $ per cent in the bulk ICM within $\sim$\Rc{2500}.
In our sample, 76 clusters (95 per cent of our sample) could fall within this range; among these, 6 clusters with the tightest constraints: A2744, the Bullet cluster, A520, MACS0717, 4C+37.11, and RXC0528.
The 4 clusters with kinetic pressure fractions below this range, those mentioned in Section \ref{sec:velocities}, have exceptionally low $P_\text{kin}/P_\text{tot}$.
Our measurements are, therefore, consistent with these simulations.

\subsection{ICM plasma physics}
\label{sec:plasma}
\subsubsection{Turbulent re-acceleration of cosmic rays}
\label{sec:radiohalos}

Other than the X-ray thermal emission, the ICM also exhibits non-thermal synchrotron emission that is bright in radio \citep[e.g.,][]{van_weeren_diffuse_2019}.
In particular, volume-filling radio halos are thought to be powered by turbulent re-acceleration of seed relativistic cosmic ray electrons (CRe) \citep[e.g.,][]{petrosian_particle_2008, brunetti_particle_2011, brunetti_challenge_2016,brunetti_relativistic_2017}.
From this model, one may expect a correlation between the observed radio power ($P_\text{1.4 GHz}$) and ICM turbulence (or kinetic energy) \citep[e.g.,][]{bonafede_lofar_2018}, likely with significant scatter due to cluster-to-cluster variation of the seed CRe population \citep{nishiwaki_statistical_2022}.

This relationship has been previously investigated (with mixed results) by \citet{eckert_connection_2017} and \citet{zhang_planck_2022}.
\citet{eckert_connection_2017} found a strong correlation between the $V_\text{1D,k}$ (at $1/k=660$ kpc) and $P_\text{1.4 GHz}$, finding a Pearson correlation coefficient of $r_{V:P} \approx 0.8$ in a sample of 25 radio halo detections {(their sample also included 26 clusters with upper limits on $P_\text{1.4 GHz}$)}.
\citet{zhang_planck_2022}, however, were unable to find any significant correlation between the same measurements in a sample of 11 halos, with $r_{V:P}\approx 0.4$ ($1/k\approx 400$ kpc). 
Both studies inferred velocities indirectly through a similar method used in this work. 
The correlation has also been explored within individual clusters.
\citet{bonafede_lofar_2018} found kinetic energy in MACS0717 $\sim 30$ per cent higher in the region with strong radio emission compared to the radio-quiet region.

\begin{figure}
    \includegraphics[width=\columnwidth]{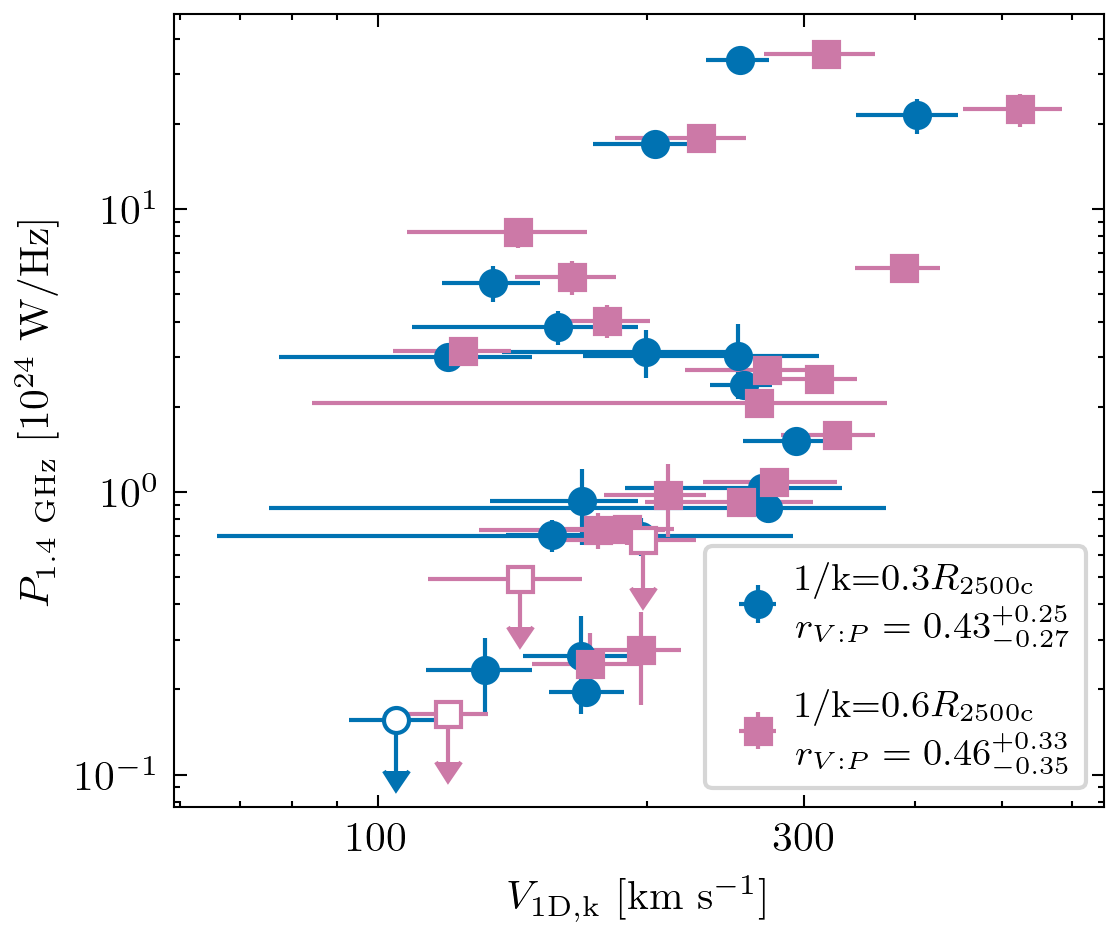}
    \vspace{-0.5cm}
    \caption{Radio halo power at 1.4 GHz (see Table \ref{tab:parameters} for references) versus characteristic velocity $V_\text{1D,k}$ at  $1/k=0.6$ (blue circles) and $0.3$ \Rc{2500} (pink squares). Unfilled squares/circles indicate upper limits on radio halo power. The Pearson correlation coefficients $r_{V:P}$ at the two different length scales are shown in legends. {Uncertainties on $r_{V:P}$ are calculated via a bootstrap and by randomly sampling measurement uncertainties on $V_\text{1D,k}$ and $P_\text{1.4 GHz}$.} {For clusters with $V_\text{1D,k}$ measured at both scales, paired pink and blue points are offset for visual clarity.}}
    \label{fig:rh_PV}
\end{figure}

In our sample, 22 clusters have detected radio halos at 1.4 GHz, in addition to 5 with upper limits (see Table \ref{tab:parameters} for details and references).
It is interesting to check the radio power - turbulence relation, especially since we are probing smaller scales that are least affected by the choice of underlying model, merging substructures, and large-scale asymmetries (Section \ref{sec:powerspec}). 
At multiple length scales, we compare the characteristic velocity to radio power in these 22 clusters, calculating $r_{V:P}$ at each scale. 
Representative distributions of $V_\text{1D,k}$ and $P_\text{1.4 GHz}$ are plotted in Figure \ref{fig:rh_PV}, where velocity is measured at 0.3 \Rc{2500} (blue circles) and 0.6 \Rc{2500} (pink squares). 
Galaxy clusters with radio halo upper limits are also plotted as unfilled points. 

{To calculate uncertainties on $r_{V:P}$, we assume the measurement uncertainties on $V_\text{1D,k}$ and $P_\text{1.4 GHz}$ are approximately Gaussian and randomly sample this distribution, finding the standard deviation on this randomized correlation coefficient.
Additionally, we estimate the sample variance of $r_{V:P}$ via a bootstrap, combining these uncertainties in quadrature.
The latter source of uncertainty is much more significant at all scales.}

\begin{figure}
    \includegraphics[width=\columnwidth]{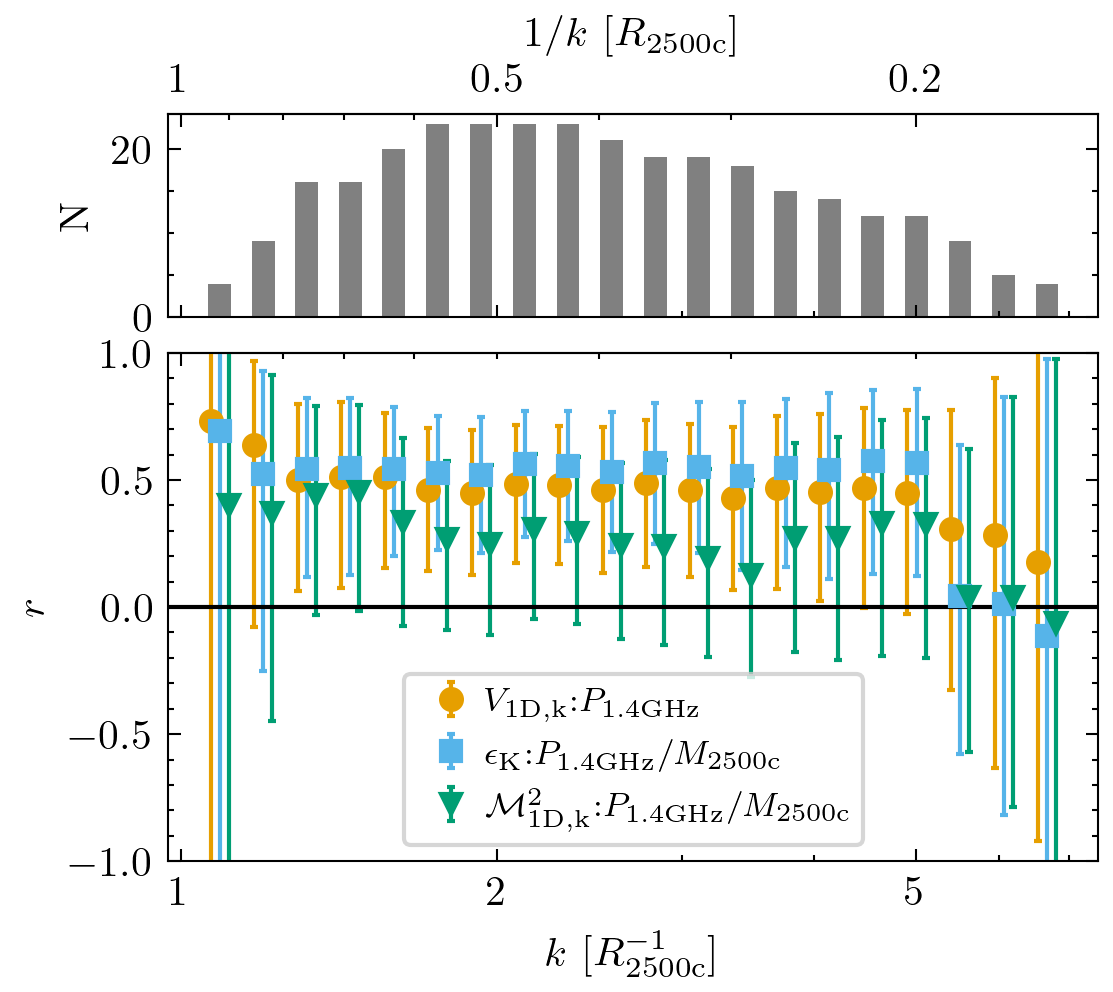}
    \vspace{-0.5cm}
    \caption{Pearson correlation coefficient $r$, with $1\sigma$ uncertainties, as a function of wavenumber/scale, comparing $V_\text{1D,k}$ to $P_\text{1.4 GHz}$ (orange), $\epsilon_\text{K}$ to $P_\text{1.4 GHz}/M_\text{2500c}$ (blue), and $\mathcal{M}_\text{1D,k}^2$ to $P_\text{1.4 GHz}/M_\text{2500c}$ (green). {Uncertainties on $r$ are calculated via a bootstrap and by randomly sampling measurement uncertainties on $V_\text{1D,k}$ and $P_\text{1.4 GHz}$.} $r=0$ is plotted as a solid black line. The number of clusters where $V_\text{1D,k}$ is measurable at each scale is plotted in the top panel. The p-value of each Pearson test is plotted in the bottom panel, with $p=0.05$ plotted as a dashed black line.}
    \label{fig:rh_pearson}
\end{figure}

{Across all length scales where velocities are measurable in most clusters, we find a mostly scale-independent correlation coefficient of $r_{V:P} \approx 0.45 \pm 0.3$.
This correlation is significantly weaker than that found by \citet{eckert_connection_2017}, and similar to that of \citet{zhang_planck_2022}.
Given our larger sample size than the latter work, we can confirm a weak positive correlation at the $p<0.05$ level between scales of $1/k \approx 0.9$ to $0.3$ \Rc{2500} (see Figure \ref{fig:rh_pearson}).}

We stress that though we see a weak positive correlation in our full sample, this correlation is driven by a single outlier cluster, MACS0717, which has the highest characteristic velocity in our sample at the scales we consider here, and the second highest radio power. 
Removing MACS0717 from the calculation results in a correlation coefficient consistent with $r_{V:P}=0$ at all wavenumbers. 
A similar but less drastic reduction in $r_{V:P}$ is seen when the Bullet cluster is removed from the sample.
These two clusters are present in the \citet{eckert_connection_2017} sample, but not in that of \citet{zhang_planck_2022}, suggesting that the strong correlation found in the former work may also be significantly driven by these same clusters.
Confirmation of this correlation, therefore, requires a larger sample of radio halo and direct velocity measurements. 

{Given the large uncertainties on $r_{V:P}$, a strong correlation cannot be excluded. 
To further investigate, we attempt to fit the $V_\text{1D,k}$---$P_\text{1.4 GHz}$ distribution to a power-law relationship.
Using the Bayesian MCMC routine \textsc{LinMix} \citep{kelly_aspects_2007}, which includes clusters with upper limits on $P_\text{1.4 GHz}$, we find that the correlation between these variables is not strong enough to recover a best-fit slope.
The slope varies significantly, between $\sim 2$ and $\sim 10$, depending on the numerical method used.
This is also the case when traditional MCMC maximum-likelihood fitting and orthogonal distance regression are used.
}

It is well-established that $P_\text{1.4 GHz}$ strongly correlates with cluster mass \citep[e.g.,][]{cassano_planck_2023, cassano_revisiting_2013}. 
This correlation could contribute to the radio power - velocity correlation. Therefore, besides radio power, we consider $P_\text{1.4 GHz}/M_\text{2500c}$, which is a proxy for the efficiency of radio emission generation. 
{This characteristic is compared to the turbulent energy dissipation rate $\epsilon_\text{K} \propto V_\text{1D,k}^3 k$ and $\mathcal{M}_\text{1D,k}^2$, which traces the ratio of kinetic to thermal energy in the ICM.}
For each of these cases, we calculate the Pearson correlation coefficient across a range of length scales from $1$ to $0.15$ \Rc{2500}, shown in the lower panel of Figure \ref{fig:rh_pearson}.
In the upper panel of this figure, we plot the number of clusters in which $V_\text{1D,k}$ is observable at each scale.
We additionally test other combinations of these variables, which do not change our conclusions.

{The strongest correlation we found is when comparing radio emission efficiency to the turbulent dissipation rate, with $r_{\epsilon: P/M} \approx 0.55 \pm 0.3$, significant at the $p<0.05$ level at length scales larger than $0.25$ \Rc{2500}.
This correlation is also significantly driven by outlier clusters, as described above.
The comparison of $P_\text{1.4 GHz}/M_\text{2500c}$ to $\mathcal{M}_\text{1D,k}^2$ is consistent with no correlation at all scales.}

Further investigation of the relationship between ICM gas motions and non-thermal emission could be improved by larger samples of radio halo detections and velocity measurements, including direct velocity measurements with XRISM that will soon become available. 
We note that $P_\text{1.4 GHz}$ is measured across the entire extent of the radio halo (with typical radii of $\sim 1$ Mpc), while our characteristic velocities are measured within \Rc{2500} ($\sim 500$ kpc). 
{The effect of this mismatch is expected to be small, as typically $\sim 90$ per cent of the total emission from radio halos is found within \Rc{2500} \citep{botteon_planck_2022}.
Additionally, the radio power measured at 1.4 GHz does not reflect the total luminosity of the halo, as there is evidence for variation in spectral slope between halos \citep{van_weeren_diffuse_2019}.}
Moreover, in some cases, significant areas around prominent substructures are removed.
Thus, the comparison between our X-ray measurements and radio powers found in the literature is not one-to-one and could be improved in future works. 

If substructures are included in the estimations of $V_\text{1D,k}$, the strength of all correlations increases on length scales smaller than $\sim 0.8$\Rc{2500}.
$r_{V:P}$ and $r_{\epsilon:P/M}$ both rise to $\sim 0.7 \pm 0.3$.
Thus, we find that improper treatment of X-ray substructures may lead to spurious correlations that do not reflect the state of the bulk ICM.

\subsubsection{Constraints on effective viscosity}
\label{sec:viscosity}

Threaded by a weak, likely tangled $\sim$ few $\mu G$ magnetic field \citep[e.g.,][]{carilli_cluster_2002}, the ICM is prone to the firehose and mirror plasma instabilities. 
These produce fluctuations, which can scatter particles, increasing their effective collisional rate and, therefore, decreasing the effective viscosity of the ICM \citep{schekochihin_turbulence_2006}. 
At the same time, the gyroradius of ions is many orders of magnitude smaller than the typical Coulomb mean free path. 
Therefore, particles are confined to magnetic field lines, making the transport of heat and momentum anisotropic with respect to the local magnetic field \citep{kunz_thermally_2011}.

Observations of prominent structures within the ICM support these predictions. 
For instance, the width of interfaces between two gas phases in cold fronts is typically narrower than the mean free path of particles, indicating suppressed diffusion, conduction and mixing across the interface \citep[e.g.,][]{markevitch_shocks_2007, werner_deep_2016, zuhone_cold_2016}. 
The presence of Kelvin-Helmholtz instabilities in cold fronts \citep[e.g.,][]{werner_deep_2016,ichinohe_azimuthally_2017, su_deep_2017,hu_merger_2021} and the morphology of stripped tails of infalling galaxies and groups \citep[e.g.,][]{kraft_stripped_2017} also support suppressed effective viscosity, typically $\lesssim 5 $ per cent, with respect to the isotropic Spitzer level \citep[$\nu_0$,][]{spitzer_physics_1962}. 
Probing the effective viscosity in the bulk ICM has recently become possible through the measurements of density fluctuation power spectra in the Coma cluster on scales comparable to the Coulomb mean free path ($\lambda_\text{mfp}$).
The shape of the spectra indicated $\nu_\text{eff}/\nu_0\lesssim 10$ per cent regardless of the level of thermal conduction \citep{zhuravleva_suppressed_2019}. 
{With our measurements, it is possible to expand the latter analysis to a sample of 16 clusters, for which \drrk{} are measured across a broad range of scales close to the mean free path.}

Using deprojected profiles of ICM temperature and density, we calculate radial profiles of $\nu_0$ \citep[see][for details]{sarazin_x-ray_1988}.
From our measurements of the scale-dependent characteristic velocity and under the assumption of Kolmogorov turbulence, we additionally find the turbulent energy dissipation rate ($\epsilon_K$) as $\epsilon_K= 2\pi\left(3 C_K /2\right)^{3/2} V_\text{1D,k}^3 k$, where $C_K\approx 1.65$ is the Kolmogorov constant \citep{zhuravleva_turbulent_2014}. 
Given $\nu_0$ and $\epsilon_K$ we find the median Kolmogorov microscale $\eta_K=\left( \nu_0^3 / \epsilon_K \right)^{1/4}$ within our region of interest. 
{The values of $\eta_K$ we find are listed in Table \ref{tab:parameters}.}

\begin{figure}
    \centering
    \includegraphics[width=\columnwidth]{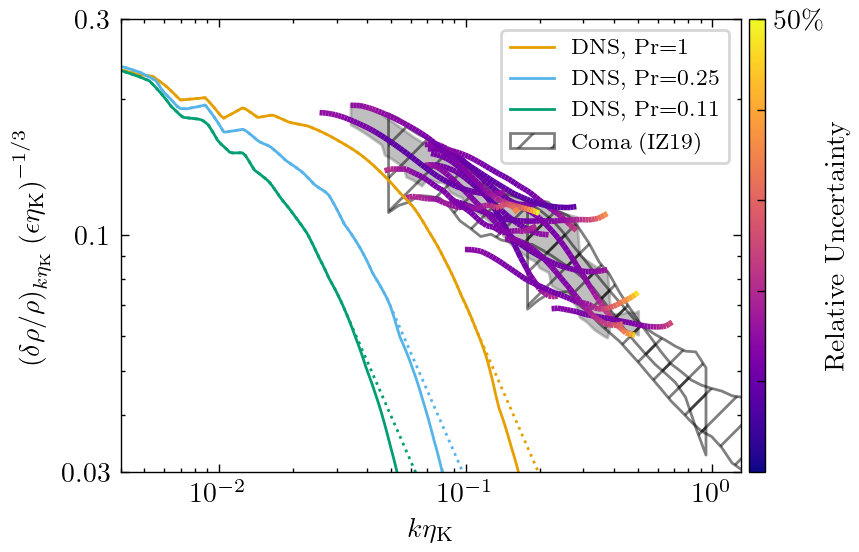}
    \vspace{-0.5cm}
    \caption{Observed rescaled $\left( \delta \rho / \rho\right)_{k\eta_K}$ in the 16 clusters in which we constrain effective viscosity. Notations are equivalent to Figure \ref{fig:allspectra}, with the sample-averaged values $\pm 1\sigma$ uncertainties shown with the grey region. The equivalent passive scalar power spectra from DNS simulations of hydrodynamic turbulence for three thermal Prandtl numbers are shown in solid lines. The dotted lines show how the exponential cutoff of the DNS spectra would be modified if the $\Delta$-variance method of calculating power spectra was used. Hatched regions show upper and lower limits on rescaled \drrk{} taken from the central and outer regions of the Coma cluster \citep{zhuravleva_suppressed_2019}, for comparison. The slope of the observed spectrum is shallower at the equivalent $k \eta_\text{K}$ than any of the DNS spectra, suggesting a suppressed effective viscosity in the ICM. }
    \label{fig:allvisc}
\end{figure}

Using these characteristics, we rescale power spectra of all clusters in our subsample\footnote{We rescale the amplitude of density fluctuations by a factor of $\left( \epsilon_\text{K} \eta_\text{K}\right)^{-1/3}$ to compare these spectra, however this normalization is arbitrary.}, see Figure \ref{fig:allvisc} (purple-orange curves), and obtain the weighted mean value (the gray region).
We also plot the earlier-measured power spectra in the Coma cluster (gray hatched regions). 
For comparison with simulations, the equivalent spectra of passive scalars from direct numerical simulations (DNS) of hydrodynamic turbulence \citep{gauding_generalised_2014} are also plotted with different colors. 
The passive scalar is characterized by the thermal Prandtl number $Pr$, the ratio of kinematic viscosity and thermal conductivity.
We use the DNS simulations for $Pr=1$, $0.25$ and $0.11$\footnote{We also compared the observed power spectra with the DNS simulations from \citet{watanabe_scalar_2007, yeung_schmidt_2002, yeung_turbulence_2014}, and \citet{ishihara_energy_2016} and confirmed that our conclusions are robust against the specific choice of the DNS simulations as long as the Reynolds number is sufficiently large.}.
We note that the Prandtl number of density fluctuations in the ICM is generally unknown, however it is expected to be less than unity \citep{kunz_plasma_2022}, thus we vary $Pr$ in this range. 
The DNS spectra predict that the passive scalar spectrum should exponentially decline/steepen on scales probed with our observations.
This steepening is not observed in any measurement of \drrk{} from this subsample, confirming earlier results based on the Coma cluster and suggesting that suppressed transport in the bulk of the gas is a general property of the ICM.

\begin{figure}
    \centering
    \includegraphics[width=\columnwidth]{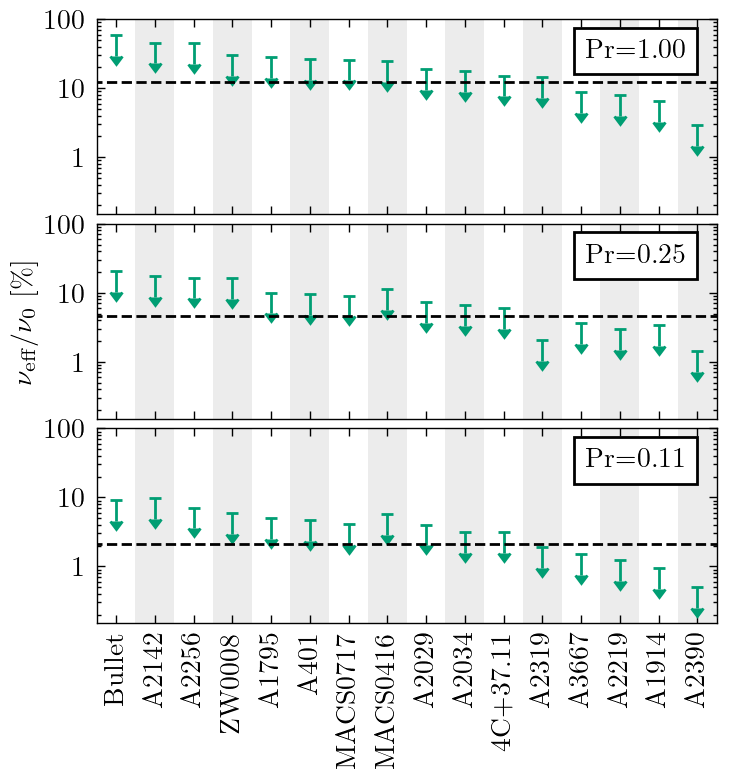}
    \vspace{-0.5cm}
    \caption{Upper limits on effective viscosity suppression, $\nu_\text{eff}/\nu_0$, relative to a Spitzer value in a subsample of 16 galaxy clusters for different thermal Prandtl numbers. Sample-averaged upper limits are plotted as dashed black lines.} 
    \vspace{-0.5cm}
    \label{fig:viscosity_constraints}
\end{figure}

Following the Coma analysis, we can use these spectra to place upper limits on $\nu_\text{eff}$. 
Suppressing $\nu_\text{eff}$ lowers $\eta_K$, essentially moving the DNS spectra to higher wavenumbers ($k \eta_\text{K}$) such that the slope of measured \drrk{} eventually matches the slope of the DNS spectra.
We use an MCMC chain with the Kolmogorov microscale and the normalization as free parameters to find the highest value of $\nu_\text{eff}/\nu_0$ that provides a reasonable fit to the observed spectrum.
These upper limits are presented for each object as green arrows in Figure \ref{fig:viscosity_constraints}.
Additionally, we find the weighted average and standard deviation of $\left( \delta \rho / \rho\right)_{k\eta_\text{K}} \left( \epsilon_\text{K} \eta_\text{K} \right)^{-1/3}$ in this subsample and find an upper limit for the average spectrum, which is plotted as a dashed black line in the same figure.

We do not find any clusters in this subsample that are consistent with Spitzer-level effective viscosity. 
With $Pr=1.0$, viscosity is suppressed by at least a factor of 2 in nearly all clusters.
With $Pr=0.11$, the suppression is at least a factor of 10 in all clusters, and a factor of 20 in all but 3 clusters. 
On average, the effective viscosity $\nu_\text{eff}/\nu_0$ is below 13 per cent in the case of $Pr=1.0$, $\lesssim 5$ per cent for $Pr=0.25$, and $\lesssim 2$ per cent for $Pr=0.11$. 
The strongest effective viscosity suppression in the sample is found in A2390, with $\nu_\text{eff}/\nu_0 \lesssim 3$ per cent for all Prandtl numbers we considered here. 
This demonstrates that, indeed, magnetic fields modify the behavior of particles in the ICM, suppressing transport properties of the bulk of the gas, and that this is a general property of ICM plasmas.

We note that the $\Delta$-variance method is designed to measure smooth, power-law power spectra. The dotted lines in Figure 
\ref{fig:allvisc} show how exponential cutoffs are modified by the $\Delta$-variance method  \citep[for details, see][]{arevalo_mexican_2012,zhuravleva_suppressed_2019}. 
Accounting for this correction does not affect the viscosity constraints shown in Figure \ref{fig:viscosity_constraints}.

Additionally, we note that we use the median measurement of the Kolmogorov microscale.
Within the regions we consider in this work, $\eta_\text{K}$ typically varies by a factor of $\sim 2$, which could increase the upper limit on $\nu_\text{eff}/\nu_0$.
If the minimum value of $\eta_\text{K}$ is used instead of the median, we still detect effective viscosity suppression in all systems except the Bullet cluster and A2142 (for the $Pr=1.0$ case).
A2390 has the largest variance, where our upper limit on $\nu_\text{eff}/\nu_0$ increases by a factor of $\sim 4.5$ if the minimum value of $\eta_\text{K}$ is used instead of the median, however $\nu_\text{eff}/\nu_0$ is still $\lesssim 13$ per cent for $Pr=1.0$, even in this conservative limit.

\section{Conclusions}

By measuring the power spectra of X-ray surface brightness fluctuations in a sample of 80 galaxy clusters observed with \chandra, we obtain the amplitude of density fluctuations and, indirectly, characteristic velocities in cluster regions dominated by merger activity, i.e., outside the cool-cores and within the \Rc{2500} radius.
Each cluster is treated individually to ensure robust measurements at the wavenumbers/length scales we present.
Observational effects (e.g., Poisson noise, PSF effects, contribution of unresolved point sources) and systematic uncertainties induced by the choice of unperturbed model and cluster substructures are carefully tested and removed.
Galaxy clusters are categorized into relaxed, intermediate, and unrelaxed merger states based on their X-ray morphology and a literature review.
Scale-dependent amplitudes of density fluctuations, \drrk{}, are averaged within each morphological category. Our main findings are summarized below.
\begin{itemize}
    \item We observe a mild increase in \drrk{} measured in the bulk ICM from relaxed to unrelaxed clusters. 
    E.g., at  $1/k=0.5$\Rc{2500}, \drrk{} $\sim 12 \pm 2 $ per cent in relaxed clusters and $\sim 19 \pm 5 $ per cent in unrelaxed ones. 
    {This suggests that fluctuations in the bulk ICM are not strongly dependent on recent merger activity.} 
    
    \item {When prominent substructures are included in the measurement, density fluctuations in merging clusters ($\sim 40$ per cent at $1/k=0.5$\Rc{2500}) are significantly higher than in relaxed clusters ($\sim 12$ per cent)}. 
    Between length scales of $\sim 0.2$ to $0.8$\Rc{2500} ({$\sim 80$ to $450$ kpc}), \drrk{} {measured including substructures} could be used as an indicator of cluster dynamic state, complementing other commonly-used methods.

    \item Power spectra of density fluctuations measured on scales from $1/k \sim 1$ to $0.2$\Rc{2500} in intermediate and unrelaxed clusters are in good agreement with theoretical predictions from cosmological simulations considered here. 

    \item On length scales smaller than $\sim 0.75$\Rc{2500}, density fluctuations measured in relaxed clusters are systematically higher {by a maximum factor of $\sim 2.5$} compared to similar clusters in cosmological simulations.
    {The effects of observational biases and numerical viscosity are likely minimal on these scales, as discussed in Section \ref{sec:simulations}. 
    The tension could be due to missing radiative and feedback physics in considered simulations. 
    Fluctuations in relaxed clusters are most sensitive to these physics compared to merging clusters, in which the amplitude of fluctuations is likely dominated by ongoing or recent mergers.
    For this reason, relaxed clusters are the most useful probes for additional, multi-scale physics in simulations compared to clusters in other dynamic states.}

    \item Estimated from power spectra of density fluctuations, clumping factors typically vary between $1.01 \lesssim C \lesssim 1.10$ within the \Rc{2500} region in our sample, with a few outliers (the Bullet and MACS0717 clusters). 
    The measured clumping factors are broadly consistent with predictions from cosmological simulations, and in good agreement with other existing measurements within a similar region. 

    \item Using a statistical relation between the amplitude of density fluctuations and velocity (with a recently-updated proportionality coefficient for clusters in different dynamical states), we find that the sample-average one-dimensional velocities are $\sim 150$ ($250$) km s$^{-1}$ at $1/k=0.5$\Rc{2500} for relaxed (unrelaxed) clusters.
    Within the uncertainties, the average power spectrum slope is consistent with a Kolmogorov cascade, although the uncertainties are large.

    \item Characteristic velocities (scale-independent) in the range of a few 100 km s$^{-1}$ are found in essentially all clusters. 
    The lowest velocities, $V_\text{1D}<200$ km s$^{-1}$, are found in A2597, A1650, A2626, A2199, and A2029.
    The highest velocities, $V_\text{1D}> 350$ km s$^{-1}$, are found in the Bullet cluster, A1916, RBS1316 and MACS0717.
    These velocities await confirmation via direct observation by \xrism.
    
    \item We confirm kinetic pressure fractions in the bulk ICM of $\lesssim 10$ per cent in 22 clusters, and $\lesssim 5$ per cent in 6 of these: A2219, A1650, PKS0745, A1835, A2199, and A2029.
    The lowest is found in A2029. 
    Note these values likely do not represent the total kinetic pressure support unless the cluster is unaffected by merging substructures (relevant to most relaxed clusters in our sample). 
    These values of non-thermal pressure support are consistent with predictions from cosmological simulations.

    \item Using a subsample of galaxy clusters that host extended radio halos, we measure a weak positive correlation between characteristic velocity and radio power between length scales of $0.9 \gtrsim 1/k \gtrsim 0.3$\Rc{2500}, consistent with \citet{zhang_planck_2022} but weaker than \citet{eckert_connection_2017}. 
    We additionally test correlations between the ``radio emission efficiency'' ($P_\text{1.4 GHz}/M_\text{2500c}$), turbulent energy dissipation rate ($\epsilon_K$), and the ratio of kinetic to thermal energy ($\mathcal{M}_\text{1D,k}^2$). 
    The strongest correlation we find is between $\epsilon_K$ and radio emission efficiency, with the Pearson $r_{\epsilon:P/M} \approx 0.55 \pm 0.3$.
    
    \item While these correlations are somewhat significant in the full sample, it is important to note that they are driven by a few outlier clusters. 
    Removing the Bullet cluster or MACS0717 from the sample significantly reduces the correlation, such that the Pearson test is consistent with zero.
    Additionally, the results can be significantly affected by the inclusion/exclusion of substructures and the choice of underlying model. 
    This demonstrates the need for upcoming direct velocity measurements with \textit{XRISM} to investigate this correlation further.

    \item Using a subsample of 16 clusters from our sample, we constrain effective viscosity in the bulk ICM by comparing the shape of the power spectra of density fluctuations with DNS simulations. 
    We find that the spectra of all 16 clusters are significantly shallower than an exponential cutoff predicted by DNS simulations on the same scales. 
    This suggests that the effective viscosity in the bulk ICM is suppressed by magnetic fields. 
    The estimated suppression is $\nu_\text{eff}/\nu_0\lesssim 13 $ per cent for $Pr=1.0$, $\lesssim 5$ per cent for $Pr=0.25$, and $\lesssim 2$ per cent for $Pr=0.11$. 
    The lowest effective viscosity is found in A2390, with $\nu_\text{eff}/\nu_0\lesssim 3$ per cent for $Pr=1.0$.
    These results are consistent with previous constraints in the Coma cluster and suggest that suppressed effective viscosity is a general property of the ICM plasmas. 
    
\end{itemize}

This work demonstrates the power of scale-dependent density fluctuations as tracers of a diverse combination of astrophysics and plasma physics.
New (\xrism) and future (\textit{Athena}, \textit{LEM}) X-ray observatories will be able to directly measure gas velocities and structure functions, calibrating the \drrk{}-$V_\text{1D,k}$ relationship we employ in this work to confirm and expand these results.
Even in this upcoming era of X-ray astronomy, the density fluctuations method will remain the observationally "cheapest" (most easily obtained) method of characterizing ICM dynamics.
Improving our understanding of the relationship between density fluctuations and gas velocities will be a subject of future numerical and observational works.

\section*{Acknowledgements}
This research has made use of data obtained from the Chandra Data Archive, and software provided by the Chandra X-ray Center (CXC) in the application of the CIAO package. 
Support for this work was provided by the National Aeronautics and Space Administration through Chandra Award Number GO1-22123A issued by the Chandra X-ray Center, which is operated by the Smithsonian Astrophysical Observatory for and on behalf of the National Aeronautics Space Administration under contract NAS8-03060. 
Additional partial support has been provided by the Alfred P. Sloan Foundation through the Sloan Research Fellowship. 
IZ was partially supported by a Clare Boothe Luce Professorship from the Henry Luce Foundation. 
{WF acknowledges support from the Smithsonian Institution, the Chandra High Resolution Camera Project through NASA contract NAS8-03060, and NASA Grant 80NSSC19K0116 and Chandra Grant GO1-22132X.} 
{RJvW acknowledges support from the ERC Starting Grant ClusterWeb 804208.}

\section*{Data Availability}
All observations used in this work (see Table \ref{tab:observations}) are or will soon be available on the \chandra public archive. 
Relevant data products will be shared upon reasonable request to the corresponding author.



\bibliographystyle{mnras}
\bibliography{heinrich23} 




\appendix
\vspace{-0.5cm}
\section{Asymmetric unperturbed models}
    \label{app:model}
\begin{figure}
    \centering
    \includegraphics[width=\columnwidth]{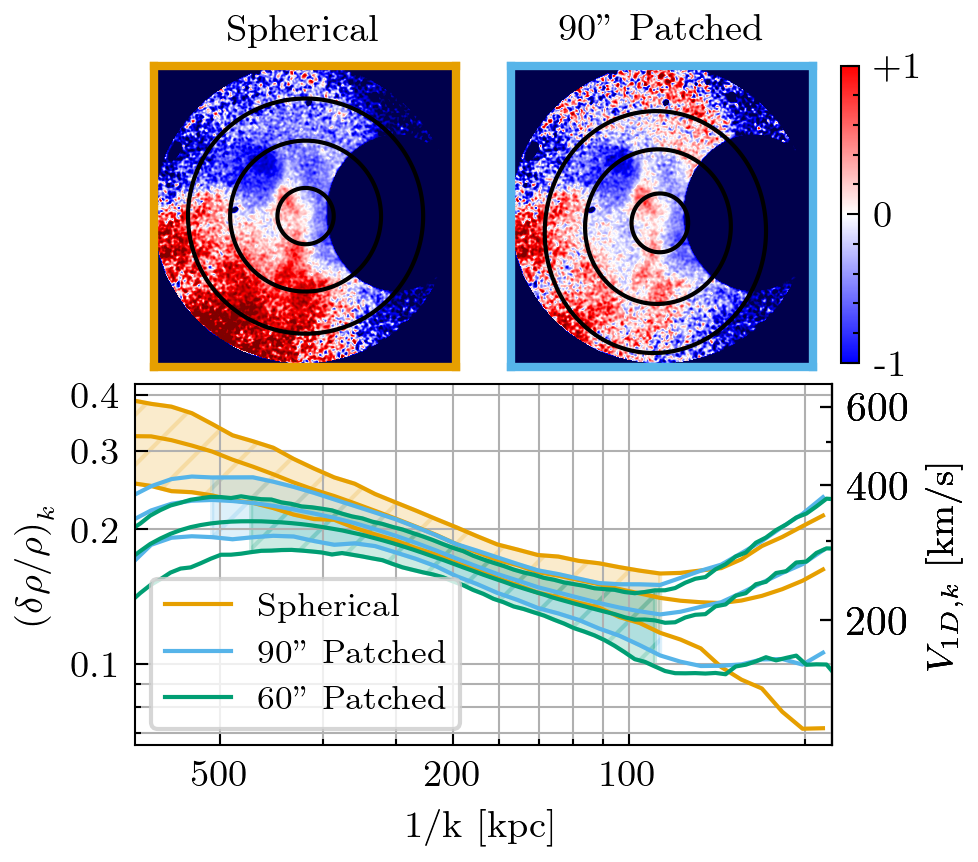}
    \vspace{-0.5cm}
    \caption{Effects of an asymmetric surface brightness model on measurements of \drrk{} in the Bullet cluster. \textbf{Top:} Residual images within \Rc{2500} of the cluster using a spherical (left) and asymmetric (right; patched at a scale of $\sim$ 400 kpc = 90") $\beta$-model for the unperturbed model. Prominent substructures have been removed as in Figure \ref{fig:masking}. \textbf{Bottom}: The orange power spectrum corresponds to the left residual image, while the blue power spectrum corresponds to the right. The green spectrum is patched on a smaller scale to ensure that measured wavenumbers are not strongly affected by the choice of patching scale. Notations are equivalent to Figure \ref{fig:masking}.}
    \vspace{-0.5cm}
    \label{fig:patching}
\end{figure}

    We measure perturbations in density relative to an underlying gas distribution to obtain gas velocities in the ICM.
    In order to do this, we must derive the best-fitting unperturbed model that reflects the global distribution of gas within the cluster.
    The nominal model we use is an elliptical single or double $\beta$-model, which has been shown to describe the ICM within our region of interest \citep{cavaliere_distribution_1978, mohr_properties_1999, kay_thermodynamic_2022}. 
    However, as there are many clusters in our sample undergoing mergers, the symmetric $\beta$-model often fails to describe the gas distribution.
    One such example is the Bullet cluster with a plane-of-sky merger that has created a strong asymmetry in the SB.
    The peak of X-ray emission from the main cluster {is due east} of the eponymous ``bullet'' cold front. However, fitting the appropriate $\beta$-model to this cluster creates a strong asymmetry, clearly seen in the top left panel of Figure \ref{fig:patching}.
    To create a model that better reflects the large-scale morphology of the cluster, we patch the unperturbed $\beta$-model.
    The residual image $R=I_\text{net}/E$, where $I_\text{net}$ is the background-subtracted X-ray counts and $E$ is the product of \chandra exposure map and the symmetric model, is smoothed with a large-scale Gaussian kernel $G$, as is the mask $M$.
    The symmetric model $S_\text{sym}$ is then modified as 
    \begin{equation}
        S_\text{asym}=S_\text{sym} \frac{G \ast R}{G \ast M}.
    \end{equation}
    The residual image based on this asymmetric model is shown in the top left panel of Figure \ref{fig:patching}.

    In practice, we test many different scales of patching, examining each resultant residual image to find which scale is sufficient to visually remove the large-scale asymmetry without over-correcting the model. 
    Our final spectra only include scales that are largely unaffected by the final choice of patching scale (see the shaded regions in the bottom panel of Figure \ref{fig:patching}).
    
\section{Supplementary figures/tables}
\begin{figure}
    \centering
    \includegraphics[width=\columnwidth]{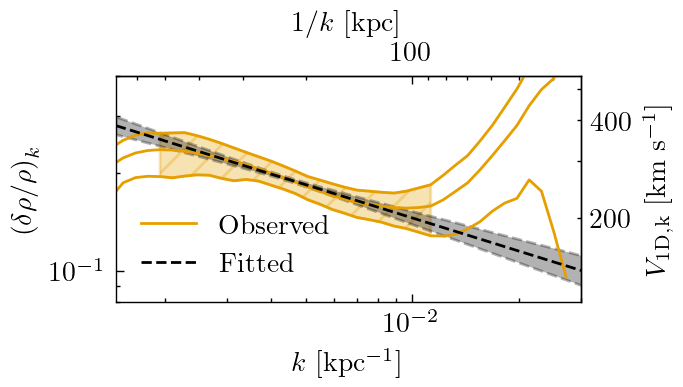}
    \caption{One example of a measured power spectrum (orange, solid lines) taken from the Bullet cluster compared to the MCMC-fitted power spectrum (black, dashed lines). $1\sigma$ uncertainties are shown in shaded regions. Notations are equivalent to Figure \ref{fig:masking}.}
    \label{fig:mcmcfit}
\end{figure}
\begin{figure*}
    \centering
    \includegraphics[width=\textwidth]{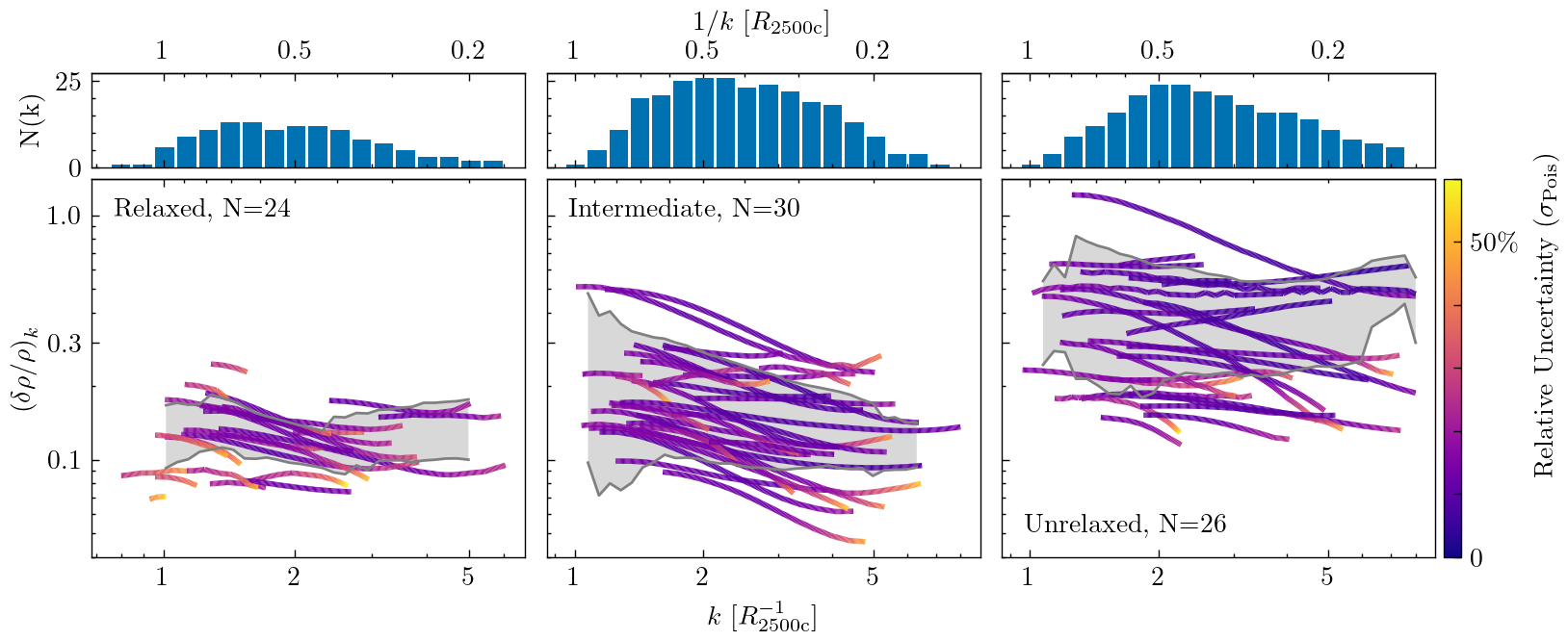}
    \caption{Equivalent to the top and middle rows of Figure \ref{fig:allspectra}, but with substructures included in the calculation of \drrk{}. }
    \label{fig:allspectra_str}
\end{figure*}

{Here we present a version of Figure \ref{fig:allspectra} with substructures included in the measurement of density fluctuations (Figure \ref{fig:allspectra_str}).
We do not calculate $V_\text{1D,k}$ with substructures included as the \drrk{}---$V_\text{1D,k}$ relationship is calibrated in the bulk ICM.
In this measurement, the difference in average amplitudes of density fluctuations between the relaxed and unrelaxed subsamples is much more prominent.

Additionally, for instructional purposes we display an example of an MCMC-fitted power spectrum for the Bullet cluster (Figure \ref{fig:mcmcfit}).
Measured \drrk{} is shown in orange, with best-fit \drrk{} in black.
This best-fit power spectrum is used to calculate upper limits on clumping factor, scale-independent velocity, and non-thermal pressure fraction listed in table \ref{tab:clumpings} and plotted in Figures \ref{fig:clumpings} and \ref{fig:machnumbers}.}

\onecolumn

\renewcommand{\arraystretch}{0.9} 
    \begin{longtable}{ p{0.08\textwidth}|p{0.05\textwidth}p{0.72\textwidth}p{0.07\textwidth} } 
    \caption{Name, redshift, \chandra ObsIDs used in this work, and cleaned exposure time (in ks) for each cluster in our sample.} \\
    \label{tab:observations}
    \input{tables/observations.tex}
    
    \end{longtable}

\renewcommand{\arraystretch}{0.1} 
    \begin{longtable}{ p{0.1\textwidth}|p{0.05\textwidth}p{0.07\textwidth}p{0.05\textwidth}p{0.06\textwidth}p{0.05\textwidth}p{0.05\textwidth}p{0.1\textwidth}p{0.05\textwidth}p{0.1\textwidth}} 
    \caption{Cluster properties: Dynamic state (Section \ref{sec:observations}), chosen center coordinates, median ICM temperature in the region of interest, cluster mass ($M_{2500c}$) and radius (\Rc{2500}), cool-core radius, total power of radio halo (if present, Section \ref{sec:radiohalos}), and estimated Kolmogorov microscale (Section \ref{sec:viscosity}). Temperatures marked with $\dagger$ are measured in this work. Temperatures marked with $\ast$ are taken from projected temperature profiles.}\\
    \label{tab:parameters}
    \input{tables/parameters.tex}
    \end{longtable}

\begin{center}
    \centering
    \renewcommand{\arraystretch}{1.25} 
    \begin{longtable}{l|llllllllll}
    \caption{Maximum and minimum length scales where \drrk{} is measurable in each cluster. Best-fit power-law parameters: slope ($\alpha$) and normalization at $k^{-1}=$\Rc{2500}.
    Upper ($_U$) and lower ($_L$) limits on $C$, $V_\text{1D}$, and kinetic pressure fractions. Lower limits (\textsc{l}) are derived from direct integration from $k=k_\text{min}$ to $k_\text{max}$, upper limits (\textsc{u}) are results of integration of the MCMC-fitted power-law from $k=1/$\Rc{2500} to $k=\infty$. We report the median value from the MCMC chain, along with uncertainties corresponding to the 16th and 84th percentile of the distribution.}\\

    \label{tab:clumpings}
    \input{tables/clumpings.tex}
    
    \end{longtable}
\end{center}

\bsp	
\label{lastpage}
\end{document}

%% file: tables/observations.tex
Cluster & Redshift & ObsIDs & Exposure\\ 
 &  &  & [ks]\\ 
\hline \\ 
\endfirsthead 
Cluster & Redshift & ObsIDs & Exposure\\ 
 &  &  & [ks]\\ 
\hline \\ 
\endhead 
CygnusA & 0.056 & 1707, 5830, 5831, 6225, 6226, 6228, 6229, 6250, 6252, 17133, 17134, 17135, 17136, 17137, 17138, 17139, 17140, 17141, 17142, 17143, 17144, 17145, 17146, 17505, 17506, 17507, 17508, 17509, 17510, 17511, 17512, 17513, 17514, 17515, 17516, 17517, 17518, 17519, 17520, 17521, 17522, 17523, 17524, 17525, 17526, 17527, 17528, 17529, 17530, 17650, 17710, 18441, 18641, 18682, 18683, 18688, 18846, 18861, 18871, 18886, 19888, 19956, 19989, 19996, 20043, 20044, 20048, 20059, 20063, 20077, 20079 &     2156\\
A1795 & 0.063 & 17225, 17226, 17228, 17230, 17231, 17232, 17610, 17611, 17612, 17615, 17618, 17622, 17624, 17625, 17628, 17629, 19871, 19873, 19874, 19875, 19939, 19972, 20646, 20647, 20648, 20650, 21833, 21834, 21835, 21838, 22832, 22834, 22835, 22836, 24603, 24605, 24606, 24607, 24608, 25049, 25672, 25676 &     1527\\
A2744 & 0.308 & 2212, 7712, 7915, 8477, 8557, 25278, 25279, 25907, 25908, 25909, 25910, 25911, 25912, 25913, 25914, 25915, 25918, 25919, 25920, 25922, 25923, 25924, 25925, 25928, 25929, 25930, 25931, 25932, 25934, 25936, 25937, 25938, 25939, 25941, 25942, 25944, 25945, 25948, 25951, 25953, 25954, 25956, 25957, 25958, 25963, 25967, 25968, 25969, 25970, 25971 &     1098\\
A2052 & 0.035 & 5807, 10477, 10478, 10479, 10480, 10879, 10914, 10915, 10916, 10917 &      615\\
A2597 & 0.085 & 6934, 7329, 19596, 19597, 19598, 20626, 20627, 20628, 20629, 20805, 20806, 20811, 20817 &      582\\
Phoenix & 0.597 & 13401, 16135, 16545, 19581, 19582, 19583, 20630, 20631, 20634, 20635, 20636, 20797 &      550\\
Bullet & 0.296 & 3184, 4984, 4985, 4986, 5355, 5356, 5357, 5358, 5361 &      524\\
A520 & 0.201 & 4215, 9424, 9425, 9426, 9430 &      512\\
MACS0947 & 0.354 & 7902, 22636, 22637, 22638, 22931, 22932, 22933, 22934, 22935, 23332, 24631, 24632, 24852, 24865 &      446\\
A2219 & 0.225 & 13988, 14355, 14356, 14431, 14451, 20588, 20589, 20785, 20951, 20952, 21966, 21967, 21968 &      444\\
A1367 & 0.022 & 17199, 17200, 17201, 17589, 17590, 17591, 17592, 18704, 18705, 18755 &      429\\
A3667 & 0.056 & 5751, 5752, 5753, 6292, 6295, 6296 &      428\\
A2319 & 0.056 & 3231, 15187, 23846, 24401, 24403, 24406, 24410, 24412, 24413, 24414, 24415, 24416, 24417, 27708, 27734, 27801, 27808, 27809 &      426\\
ZW0008 & 0.104 & 15318, 17204, 17205, 18242, 18243, 18244, 19901, 19902, 19905, 19916 &      412\\
A2142 & 0.089 & 5005, 7692, 15186, 16564, 16565, 17168, 17169, 17492 &      395\\
A1240 & 0.160 & 4961, 22646, 22647, 22720, 22965, 23060, 23061, 23145, 23154, 23155, 23165, 23176, 23180, 23187 &      351\\
MACS1149 & 0.544 & 3589, 16238, 16239, 16306, 16582, 17595, 17596 &      341\\
MACS0416 & 0.397 & 10446, 16236, 16237, 16304, 16523, 17313 &      324\\
A773 & 0.217 & 5006, 21540, 21541, 21542, 21543, 22043, 22044, 22045, 22051, 22052 &      276\\
A780 & 0.054 & 4969, 4970, 19570, 19571, 19997, 20007, 20968, 20969, 20970, 20971 &      256\\
A2034 & 0.113 & 2204, 7695, 12885, 12886, 13192, 13193 &      255\\
A98N & 0.104 & 11876, 11877, 12185, 21534, 21535, 21856, 21857, 21880, 21893, 21894, 21895 &      248\\
A1664 & 0.128 & 1648, 7901, 17172, 17173, 17557, 17568 &      245\\
A1650 & 0.084 & 4178, 5822, 5823, 6356, 6357, 6358, 7242, 7691, 10424, 10425, 10426, 10427 &      241\\
A133 & 0.057 & 2203, 3183, 3710, 9897, 13518 &      241\\
MACS0717 & 0.546 & 1655, 4200, 16235, 16305 &      239\\
3C186 & 1.063 & 3098, 9407, 9408, 9774, 9775 &      224\\
ZW2341 & 0.283 & 5786, 17170, 17490, 18702, 18703 &      223\\
MACS1347 & 0.447 & 3592, 13516, 13999, 14407 &      208\\
A644 & 0.070 & 2211, 10420, 10421, 10422, 23853, 24311, 24312, 24313, 24314, 24315, 24918 &      207\\
A3847 & 0.153 & 11506, 15091, 15092, 15643 &      206\\
AS0295 & 0.300 & 12260, 16127, 16282, 16524, 16525, 16526 &      204\\
A1758N & 0.279 & 2213, 7710, 13997, 15538, 15540 &      202\\
A85 & 0.055 & 15173, 15174, 16263, 16264 &      202\\
A3411 & 0.162 & 15316, 17193, 17496, 17497, 17583, 17584, 17585 &      201\\
MACS1206 & 0.440 & 3277, 20544, 20929, 21078, 21079, 21081 &      199\\
A401 & 0.074 & 10416, 10417, 10418, 10419, 14024, 22635, 22882 &      195\\
PKS0745 & 0.103 & 2427, 6103, 7694, 12881, 19572, 19573, 19574, 19575 &      190\\
A1835 & 0.253 & 6880, 6881, 7370 &      190\\
SPT2043 & 0.723 & 13478, 18240 &      187\\
A1689 & 0.183 & 1663, 5004, 6930, 7289, 7701 &      186\\
A2256 & 0.058 & 2419, 16129, 16514, 16515, 16516 &      184\\
A586 & 0.171 & 11723, 18278, 18279, 19961, 19962, 20003, 20004 &      174\\
ZW0104 & 0.119 & 15152, 21557, 22879, 23065, 23210 &      162\\
RXC1504 & 0.215 & 4935, 5793, 17197, 17669, 17670 &      161\\
MACS1621 & 0.461 & 3254, 3594, 6109, 6172, 7720, 9379, 10785 &      160\\
MACS0025 & 0.586 & 3251, 5010, 10413, 10786, 10797 &      157\\
3C438 & 0.290 & 3967, 12879, 13218 &      156\\
A478 & 0.059 & 1669, 6102, 6928, 6929, 7217, 7218, 7222, 7231, 7232, 7233, 7234, 7235 &      151\\
A4067 & 0.099 & 20530, 22123, 22124, 22125 &      149\\
MACS1226 & 0.436 & 3590, 12878 &      147\\
A1413 & 0.143 & 1661, 5002, 5003, 7696, 12194, 12195, 12196, 13128 &      146\\
A3112 & 0.075 & 2216, 2516, 6972, 7323, 7324, 13135 &      144\\
A665 & 0.182 & 3586, 7700, 12286, 13201, 15145, 15146, 15147, 15148 &      143\\
A1351 & 0.322 & 15136, 23845, 24296, 24297, 24298, 25101 &      140\\
A1914 & 0.171 & 3593, 18252, 20023, 20024, 20025, 20026 &      139\\
A2495 & 0.077 & 12876, 23849, 24277, 24278, 24279, 24650, 24659 &      137\\
A2626 & 0.055 & 3192, 16136 &      134\\
A2443 & 0.107 & 12257, 15169, 20625, 21437 &      133\\
MACS1423 & 0.545 & 1657, 4195 &      133\\
ZW0949 & 0.214 & 3195, 7706, 12903 &      126\\
AS1063 & 0.347 & 4966, 18611, 18818 &      124\\
A2199 & 0.030 & 10748, 10803, 10804, 10805 &      119\\
MACS2243 & 0.447 & 3260, 19749, 20129 &      119\\
MACS1931 & 0.352 & 3282, 9382 &      112\\
RXC1532 & 0.345 & 1649, 1665, 14009 &      108\\
A2204 & 0.151 & 6104, 7940, 12895, 12896, 12897, 12898 &      107\\
4C+37.11 & 0.055 & 12704, 16120 &      104\\
RXC0528 & 0.263 & 4994, 15177, 15658 &      104\\
ZW1454 & 0.258 & 4192, 7709 &       99\\
A2029 & 0.078 & 4977, 6101, 10434, 10435, 10436, 10437 &       98\\
A1300 & 0.307 & 3276, 15300, 19775, 20988, 20989, 20990 &       93\\
A2390 & 0.231 & 4193 &       84\\
RXC0439 & 0.230 & 1449, 1506, 3583, 3905, 3906 &       79\\
A2537 & 0.297 & 4962, 9372 &       74\\
MACS2140 & 0.313 & 4974, 5250 &       63\\
ZW1518 & 0.045 & 900 &       57\\
A2063 & 0.072 & 4187, 5795, 6262, 6263 &       50\\
A2151 & 0.037 & 17038, 18171, 19592, 20086, 20087 &       38\\
MACS2228 & 0.412 & 3285 &       20

%% file: tables/parameters.tex
Cluster & State & (RA,Dec) & T & $M_\mathrm{2500c}$ & $R_\mathrm{2500c}$ & $R_\mathrm{cool}$ & $P_\mathrm{1.4 \ GHz}$ & $\eta_\mathrm{K}$ & References\\ 
 &  & [deg] & [keV] & [$10^{14} M_\odot$] & [kpc] & [kpc] & [$10^{24}$ W/Hz] & [kpc] & \\ 
\hline \\ 
\endfirsthead 
Cluster & State & (RA,Dec) & T & $M_\mathrm{2500c}$ & $R_\mathrm{2500c}$ & $R_\mathrm{cool}$ & $P_\mathrm{1.4 \ GHz}$ & $\eta_\mathrm{K}$ & References\\ 
 &  & [deg] & [keV] & [$10^{14} M_\odot$] & [kpc] & [kpc] & [$10^{24}$ W/Hz] & [kpc] & \\ 
\hline \\ 
\endhead 
CygnusA & I & (299.8682, 40.7339) & 8.1$^\dagger$ & 1.36 & 461 & 86 & -- & -- & \makecell[tl]{\citealp{comis_x-ray_2011} \\ \citealp{hudson_what_2010}}\\
A1795 & I & (207.2195, 26.5906) & 6.8$^\dagger$ & 3.80 & 597 & 189 & -- & $28^{+21}_{-11}$ & \makecell[tl]{\citealp{comis_x-ray_2011} \\ \citealp{geng_chandra_2022}}\\
A2744 & U & (3.5787, -30.3906) & 10.0$^\ast$ & 2.56 & 647 & -- & $17.4\pm0.9$ & -- & \makecell[tl]{\citealp{cavagnolo_bandpass_2008} \\ \citealp{cavagnolo_intracluster_2009} \\ \citealp{million_chandra_2009} \\ \citealp{pearce_vla_2017}}\\
A2052 & I & (229.1851, 7.0216) & 3.2$^\dagger$ & 0.42 & 307 & 123 & -- & -- & \makecell[tl]{\citealp{cavagnolo_intracluster_2009} \\ \citealp{comis_x-ray_2011}}\\
A2597 & R & (351.3325, -12.1248) & 4.2$^\dagger$ & 0.99 & 388 & 86 & -- & -- & \makecell[tl]{\citealp{mantz_weighing_2016} \\ \citealp{morris_xmm-newton_2005}}\\
Phoenix & R & (356.1825, -42.7201) & 14.0 & 5.75 & 592 & 300 & -- & -- & \makecell[tl]{\citealp{kitayama_deeply_2020} \\ \citealp{mantz_cosmology_2014} \\ \citealp{mcdonald_anatomy_2019}}\\
Bullet & U & (104.6211, -55.9417) & 10.0$^\ast$ & 7.41 & 688 & -- & $34.4\pm1.1$ & $13^{+36}_{-5}$ & \makecell[tl]{\citealp{cavagnolo_bandpass_2008} \\ \citealp{cavagnolo_intracluster_2009} \\ \citealp{million_chandra_2009} \\ \citealp{sikhosana_meerkats_2022}}\\
A520 & U & (73.5437, 2.9252) & 9.0 & 1.76 & 496 & -- & $2.45\pm0.18$ & -- & \makecell[tl]{\citealp{cassano_revisiting_2013} \\ \citealp{cavagnolo_intracluster_2009} \\ \citealp{hicks_multiwavelength_2006} \\ \citealp{zhu_chandra_2016}}\\
MACS0947 & R & (146.8039, 76.3869) & 9.0 & 4.04 & 480 & 200 & -- & -- & \makecell[tl]{\citealp{cavagnolo_powerful_2011} \\ \citealp{comis_x-ray_2011} \\ \citealp{ubertosi_multiple_2023}}\\
A2219 & I & (250.0850, 46.7082) & 13.0 & 4.54 & 463 & 45 & $5.63\pm0.80$ & $37^{+7}_{-14}$ & \makecell[tl]{\citealp{canning_series_2017} \\ \citealp{cassano_revisiting_2013} \\ \citealp{cavagnolo_bandpass_2008} \\ \citealp{cavagnolo_intracluster_2009}}\\
A1367 & U & (176.1951, 19.7155) & 3.3$^\ast$ & 0.65 & 311 & -- & $<0.16$ & -- & \makecell[tl]{\citealp{farnsworth_discovery_2013} \\ \citealp{hudson_what_2010} \\ \citealp{phipps_expanding_2019} \\ \citealp{sanderson_statistically-selected_2006}}\\
A3667 & U & (303.1458, -56.8430) & 7.0$^\ast$ & 1.02 & 408 & -- & -- & $19^{+3}_{-1}$ & \makecell[tl]{\citealp{akamatsu_properties_2012} \\ \citealp{comis_x-ray_2011} \\ \citealp{rossetti_cool_2010}}\\
A2319 & I & (290.2997, 43.9470) & 9.5$^\dagger$ & 2.34 & 538 & 120 & -- & $35^{+18}_{-14}$ & \makecell[tl]{\citealp{cavagnolo_intracluster_2009} \\ \citealp{ettori_hydrostatic_2019}}\\
ZW0008 & U & (2.9544, 52.5428) & 6.5$^\ast$ & 0.69 & 353 & -- & -- & $11^{+5}_{-4}$ & \makecell[tl]{\citealp{di_gennaro_evidence_2019} \\ \citealp{golovich_mc2_2017}}\\
A2142 & I & (239.5856, 27.2293) & 8.7$^\dagger$ & 2.25 & 528 & 115 & -- & $24^{+22}_{-10}$ & \makecell[tl]{\citealp{cavagnolo_intracluster_2009} \\ \citealp{comis_x-ray_2011}}\\
A1240 & U & (170.9012, 43.1125) & 4.5$^\dagger$$^\ast$ & 0.90 & 374 & -- & -- & -- & \makecell[tl]{\citealp{cavagnolo_intracluster_2009} \\ \citealp{cho_multiwavelength_2022}}\\
MACS1149 & U & (177.3962, 22.4030) & 10.0$^\ast$ & 1.50 & 397 & -- & $3.1\pm0.9$ & -- & \makecell[tl]{\citealp{bonafede_discovery_2012} \\ \citealp{bruno_lofar_2021} \\ \citealp{cavagnolo_intracluster_2009} \\ \citealp{comis_x-ray_2011}}\\
MACS0416 & I & (64.0389, -24.0676) & 10.5 & 2.83 & 508 & -- & $1.06\pm0.09$ & $22^{+3}_{-2}$ & \makecell[tl]{\citealp{diego_free_2015} \\ \citealp{ogrean_frontier_2015} \\ \citealp{shitanishi_thermodynamic_2018}}\\
A773 & U & (139.4716, 51.7279) & 7.7 & 3.12 & 615 & -- & $0.90\pm0.10$ & -- & \makecell[tl]{\citealp{botteon_planck_2022} \\ \citealp{cavagnolo_bandpass_2008} \\ \citealp{cavagnolo_intracluster_2009} \\ \citealp{zhu_chandra_2016}}\\
A780 & R & (139.5243, -12.0959) & 3.8$^\ast$ & 0.80 & 376 & 150 & -- & -- & \makecell[tl]{\citealp{cavagnolo_intracluster_2009} \\ \citealp{mantz_weighing_2016} \\ \citealp{sato_suzaku_2012}}\\
A2034 & U & (227.5490, 33.5094) & 8.0$^\ast$ & 1.94 & 498 & -- & $0.72\pm0.09$ & $25^{+26}_{-4}$ & \makecell[tl]{\citealp{botteon_planck_2022} \\ \citealp{cavagnolo_intracluster_2009} \\ \citealp{comis_x-ray_2011} \\ \citealp{kempner_chandra_2003}}\\
A98N & I & (11.6034, 20.6216) & 3.0 & 0.38 & 270 & 100 & -- & -- & \makecell[tl]{\citealp{alvarez_suzaku_2022}}\\
A1664 & I & (195.9266, -24.2458) & 4.5 & 1.91 & 506 & 100 & -- & -- & \makecell[tl]{\citealp{calzadilla_revealing_2019} \\ \citealp{comis_x-ray_2011} \\ \citealp{kirkpatrick_chandra_2009}}\\
A1650 & I & (194.6728, -1.7615) & 6.2$^\dagger$ & 0.93 & 397 & 86 & -- & -- & \makecell[tl]{\citealp{cavagnolo_intracluster_2009} \\ \citealp{comis_x-ray_2011}}\\
A133 & R & (15.6743, -21.8798) & 4.7$^\dagger$ & 0.71 & 365 & 92 & -- & -- & \makecell[tl]{\citealp{cavagnolo_intracluster_2009} \\ \citealp{comis_x-ray_2011}}\\
MACS0717 & U & (109.3811, 37.7576) & 18.0 & 3.24 & 507 & -- & $22\pm3$ & $21^{+8}_{-8}$ & \makecell[tl]{\citealp{cavagnolo_intracluster_2009} \\ \citealp{comis_x-ray_2011} \\ \citealp{rajpurohit_physical_2021} \\ \citealp{shitanishi_thermodynamic_2018}}\\
3C186 & R & (116.0730, 37.8874) & 7.5 & 1.79 & 320 & 120 & -- & -- & \makecell[tl]{\citealp{mantz_cosmology_2014} \\ \citealp{mantz_cosmology_2016} \\ \citealp{siemiginowska_high-redshift_2010}}\\
ZW2341 & U & (355.9203, 0.3077) & 6.0$^\ast$ & 1.66 & 444 & -- & $0.717\pm0.108$ & -- & \makecell[tl]{\citealp{akamatsu_systematic_2013} \\ \citealp{parekh_moss_2021} \\ \citealp{planck_collaboration_planck_2015} \\ \citealp{zhang_deep_2021}}\\
MACS1347 & I & (206.8772, -11.7525) & 16.0 & 10.70 & 783 & 172 & -- & -- & \makecell[tl]{\citealp{cavagnolo_intracluster_2009} \\ \citealp{comis_x-ray_2011} \\ \citealp{shitanishi_thermodynamic_2018}}\\
A644 & I & (124.3526, -7.5076) & 8.3 & 2.09 & 517 & 132 & -- & -- & \makecell[tl]{\citealp{cavagnolo_intracluster_2009} \\ \citealp{zhu_chandra_2016}}\\
A3847 & R & (333.6061, -17.0264) & 3.5$^\ast$ & 1.03 & 325 & 173 & -- & -- & \makecell[tl]{\citealp{sereno_comalit-v_2017} \\ \citealp{vagshette_detection_2017}}\\
AS0295 & U & (41.3720, -53.0372) & 8.5$^\dagger$ & 2.48 & 504 & -- & -- & -- & \makecell[tl]{\citealp{pascut_chandra_2019}}\\
A1758N & U & (203.1831, 50.5434) & 9.8$^\ast$ & 2.56 & 513 & -- & $3.93\pm0.53$ & -- & \makecell[tl]{\citealp{botteon_lofar_2018} \\ \citealp{botteon_planck_2022} \\ \citealp{cavagnolo_intracluster_2009} \\ \citealp{landry_chandra_2013}}\\
A85 & I & (10.4604, -9.3034) & 6.0$^\dagger$ & 1.29 & 443 & 107 & -- & -- & \makecell[tl]{\citealp{cavagnolo_intracluster_2009} \\ \citealp{comis_x-ray_2011}}\\
A3411 & U & (130.4922, -17.4981) & 7.0 & 2.31 & 520 & -- & $0.27\pm0.1$ & -- & \makecell[tl]{\citealp{andrade-santos_chandra_2019} \\ \citealp{cuciti_radio_2021} \\ \citealp{zhang_x-ray_2020}}\\
MACS1206 & I & (181.5518, -8.8023) & 14.0 & 2.84 & 511 & 162 & -- & -- & \makecell[tl]{\citealp{cavagnolo_intracluster_2009} \\ \citealp{comis_x-ray_2011} \\ \citealp{shitanishi_thermodynamic_2018}}\\
A401 & I & (44.7393, 13.5803) & 8.1$^\dagger$ & 2.90 & 596 & 107 & $0.2\pm0.02$ & $24^{+19}_{-6}$ & \makecell[tl]{\citealp{cavagnolo_intracluster_2009} \\ \citealp{comis_x-ray_2011} \\ \citealp{murgia_double_2010}}\\
PKS0745 & I & (116.8805, -19.2944) & 9.0$^\dagger$ & 2.98 & 579 & 200 & -- & -- & \makecell[tl]{\citealp{cavagnolo_intracluster_2009} \\ \citealp{comis_x-ray_2011}}\\
A1835 & I & (210.2580, 2.8788) & 9.2$^\dagger$ & 3.50 & 591 & 168 & -- & -- & \makecell[tl]{\citealp{cavagnolo_intracluster_2009} \\ \citealp{comis_x-ray_2011}}\\
SPT2043 & R & (310.8233, -50.5923) & 6.5 & 1.50 & 360 & 100 & -- & -- & \makecell[tl]{\citealp{mcdonald_detailed_2019} \\ \citealp{sanders_hydrostatic_2018}}\\
A1689 & I & (197.8731, -1.3416) & 9.1$^\dagger$ & 6.22 & 710 & 198 & $0.95\pm.28$ & -- & \makecell[tl]{\citealp{cavagnolo_intracluster_2009} \\ \citealp{cuciti_radio_2021} \\ \citealp{martino_locuss_2014}}\\
A2256 & U & (255.9645, 78.6486) & 7.5$^\ast$ & 1.90 & 509 & -- & $0.24\pm0.07$ & $15^{+12}_{-3}$ & \makecell[tl]{\citealp{comis_x-ray_2011} \\ \citealp{ge_chandra_2020} \\ \citealp{hudson_what_2010} \\ \citealp{rajpurohit_deep_2023}}\\
A586 & R & (113.0844, 31.6322) & 6.8$^\ast$ & 1.70 & 309 & 151 & -- & -- & \makecell[tl]{\citealp{cavagnolo_intracluster_2009} \\ \citealp{comis_x-ray_2011} \\ \citealp{landry_chandra_2013}}\\
ZW0104 & U & (16.9504, 54.1360) & 7.5$^\ast$ & 2.39 & 531 & -- & -- & -- & \makecell[tl]{\citealp{randall_multi-wavelength_2016}}\\
RXC1504 & R & (226.0313, -2.8044) & 10.8$^\dagger$ & 3.83 & 607 & 165 & -- & -- & \makecell[tl]{\citealp{cavagnolo_intracluster_2009} \\ \citealp{comis_x-ray_2011}}\\
MACS1621 & R & (245.3521, 38.1691) & 8.5 & 3.06 & 507 & 160 & -- & -- & \makecell[tl]{\citealp{cavagnolo_intracluster_2009} \\ \citealp{mantz_cosmology_2014} \\ \citealp{mantz_cosmology_2016}}\\
MACS0025 & U & (6.3733, -12.3777) & 9.5 & 2.16 & 431 & -- & -- & -- & \makecell[tl]{\citealp{pascut_chandra_2015} \\ \citealp{sayers_imaging_2019} \\ \citealp{shitanishi_thermodynamic_2018}}\\
3C438 & U & (328.9702, 37.9999) & 8.5$^\ast$ & 2.35 & 497 & -- & -- & -- & \makecell[tl]{\citealp{botteon_shocks_2018} \\ \citealp{emery_spectacular_2017}}\\
A478 & R & (63.3567, 10.4669) & 8.1$^\dagger$ & 2.80 & 553 & 171 & -- & -- & \makecell[tl]{\citealp{cavagnolo_intracluster_2009} \\ \citealp{comis_x-ray_2011}}\\
A4067 & U & (359.7048, -60.6198) & 3.5$^\ast$ & 0.32 & 345 & -- & -- & -- & \makecell[tl]{\citealp{cavagnolo_intracluster_2009} \\ \citealp{chon_witnessing_2015}}\\
MACS1226 & I & (186.7130, 21.8316) & 5.5$^\dagger$ & 1.14 & 370 & -- & -- & -- & \makecell[tl]{\citealp{de_propris_deep_2013} \\ \citealp{ehlert_co-evolution_2013}}\\
A1413 & R & (178.8246, 23.4057) & 7.5$^\dagger$ & 1.60 & 465 & 142 & -- & -- & \makecell[tl]{\citealp{cavagnolo_intracluster_2009} \\ \citealp{comis_x-ray_2011}}\\
A3112 & R & (49.4923, -44.2368) & 5.3$^\dagger$ & 0.91 & 392 & 95 & -- & -- & \makecell[tl]{\citealp{cavagnolo_intracluster_2009} \\ \citealp{comis_x-ray_2011}}\\
A665 & U & (127.7293, 65.8564) & 7.5$^\ast$ & 1.31 & 433 & -- & $3.08\pm0.33$ & -- & \makecell[tl]{\citealp{botteon_planck_2022} \\ \citealp{cavagnolo_intracluster_2009} \\ \citealp{comis_x-ray_2011} \\ \citealp{dasadia_strong_2016}}\\
A1351 & U & (175.6025, 58.5272) & 9.9$^\dagger$ & 2.25 & 484 & -- & $8.06\pm1.01$ & -- & \makecell[tl]{\citealp{botteon_planck_2022} \\ \citealp{cuciti_radio_2021}}\\
A1914 & U & (216.5016, 37.8257) & 8.8$^\dagger$ & 5.80 & 683 & -- & $1.55\pm0.11$ & $35^{+34}_{-7}$ & \makecell[tl]{\citealp{cavagnolo_intracluster_2009} \\ \citealp{comis_x-ray_2011} \\ \citealp{mandal_ultra-steep_2019}}\\
A2495 & R & (342.5828, 10.9043) & 4.4$^\dagger$ & 0.90 & 294 & 70 & -- & -- & \makecell[tl]{}\\
A2626 & I & (354.1266, 21.1468) & 3.4$^\dagger$ & 0.43 & 308 & 66 & -- & -- & \makecell[tl]{\citealp{cavagnolo_intracluster_2009} \\ \citealp{comis_x-ray_2011}}\\
A2443 & U & (336.5180, 17.3780) & 8.5$^\ast$ & 1.22 & 426 & -- & -- & -- & \makecell[tl]{\citealp{clarke_chandra_2013} \\ \citealp{golovich_merging_2019} \\ \citealp{planck_collaboration_planck_2015}}\\
MACS1423 & R & (215.9495, 24.0783) & 9.0 & 4.80 & 568 & 160 & -- & -- & \makecell[tl]{\citealp{cavagnolo_intracluster_2009} \\ \citealp{comis_x-ray_2011} \\ \citealp{shitanishi_thermodynamic_2018}}\\
ZW0949 & R & (148.2050, 51.8847) & 5.5 & 1.29 & 426 & 85 & $<0.48$ & -- & \makecell[tl]{\citealp{bruno_planck_2023} \\ \citealp{cavagnolo_intracluster_2009} \\ \citealp{comis_x-ray_2011} \\ \citealp{mantz_cosmology_2016}}\\
AS1063 & I & (342.1882, -44.5285) & 18.0 & 5.50 & 662 & 197 & $2.63\pm0.18$ & -- & \makecell[tl]{\citealp{cavagnolo_intracluster_2009} \\ \citealp{comis_x-ray_2011} \\ \citealp{shitanishi_thermodynamic_2018} \\ \citealp{xie_discovery_2020}}\\
A2199 & I & (247.1591, 39.5505) & 4.4$^\dagger$ & 0.27 & 266 & 60 & -- & -- & \makecell[tl]{\citealp{comis_x-ray_2011} \\ \citealp{kawano_chandra_2003}}\\
MACS2243 & U & (340.8405, -9.5947) & 7.0$^\ast$ & 5.52 & 622 & -- & $3.2\pm0.6$ & -- & \makecell[tl]{\citealp{cantwell_newly-discovered_2016} \\ \citealp{cavagnolo_intracluster_2009} \\ \citealp{ehlert_co-evolution_2013} \\ \citealp{parekh_study_2017}}\\
MACS1931 & I & (292.9567, -26.5761) & 10.0 & 5.20 & 645 & 172 & -- & -- & \makecell[tl]{\citealp{cavagnolo_intracluster_2009} \\ \citealp{comis_x-ray_2011} \\ \citealp{shitanishi_thermodynamic_2018}}\\
RXC1532 & R & (233.2242, 30.3494) & 6.5 & 2.90 & 499 & 160 & $<0.66$ & -- & \makecell[tl]{\citealp{cassano_revisiting_2013} \\ \citealp{cavagnolo_intracluster_2009} \\ \citealp{comis_x-ray_2011} \\ \citealp{shitanishi_thermodynamic_2018}}\\
A2204 & I & (248.1958, 5.5752) & 5.9$^\dagger$ & 9.20 & 830 & 131 & -- & -- & \makecell[tl]{\citealp{cavagnolo_intracluster_2009} \\ \citealp{comis_x-ray_2011}}\\
4C+37.11 & R & (61.4548, 38.0589) & 5.5$^\ast$ & 0.82 & 380 & 97 & -- & $44^{+15}_{-7}$ & \makecell[tl]{\citealp{andrade-santos_binary_2016}}\\
RXC0528 & I & (82.2197, -39.4733) & 9.1 & 2.50 & 520 & 96 & -- & -- & \makecell[tl]{\citealp{cavagnolo_intracluster_2009} \\ \citealp{comis_x-ray_2011} \\ \citealp{zhu_chandra_2016}}\\
ZW1454 & I & (224.3119, 22.3410) & 5.0$^\ast$ & 1.80 & 460 & 180 & -- & -- & \makecell[tl]{\citealp{cavagnolo_intracluster_2009} \\ \citealp{landry_chandra_2013} \\ \citealp{planck_collaboration_planck_2013}}\\
A2029 & I & (227.7337, 5.7448) & 9.2$^\dagger$ & 1.52 & 455 & 180 & -- & $47^{+25}_{-14}$ & \makecell[tl]{\citealp{cavagnolo_intracluster_2009} \\ \citealp{comis_x-ray_2011}}\\
A1300 & U & (172.9833, -19.9253) & 10.5$^\ast$ & 3.28 & 552 & -- & $6.055\pm0.56$ & -- & \makecell[tl]{\citealp{cuciti_radio_2021} \\ \citealp{kale_extended_2015} \\ \citealp{parekh_study_2017}}\\
A2390 & I & (328.4036, 17.6951) & 12.4$^\dagger$ & 8.43 & 701 & 122 & -- & $87^{+112}_{-59}$ & \makecell[tl]{\citealp{cavagnolo_intracluster_2009} \\ \citealp{hicks_multiwavelength_2006}}\\
RXC0439 & R & (69.7521, 7.2679) & 6.5 & 1.84 & 468 & 100 & $<0.43$ & -- & \makecell[tl]{\citealp{cavagnolo_intracluster_2009} \\ \citealp{cuciti_radio_2021} \\ \citealp{kale_extended_2013} \\ \citealp{mantz_cosmology_2016}}\\
A2537 & R & (347.0919, -2.1910) & 8.8 & 1.37 & 414 & 104 & $<0.48$ & -- & \makecell[tl]{\citealp{cavagnolo_intracluster_2009} \\ \citealp{kale_extended_2015} \\ \citealp{zhu_chandra_2016}}\\
MACS2140 & R & (325.0633, -23.6611) & 10.0 & 4.20 & 576 & 136 & -- & -- & \makecell[tl]{\citealp{cavagnolo_intracluster_2009} \\ \citealp{comis_x-ray_2011} \\ \citealp{shitanishi_thermodynamic_2018}}\\
ZW1518 & R & (230.4660, 7.7060) & 4.0$^\dagger$ & 0.67 & 357 & 122 & -- & -- & \makecell[tl]{\citealp{cavagnolo_intracluster_2009} \\ \citealp{comis_x-ray_2011}}\\
A2063 & R & (230.7704, 8.6072) & 4.0$^\dagger$ & 0.90 & 395 & 96 & -- & -- & \makecell[tl]{\citealp{cavagnolo_intracluster_2009} \\ \citealp{comis_x-ray_2011}}\\
A2151 & I & (241.1490, 17.7217) & 2.2$^\dagger$ & 0.19 & 237 & 29 & -- & -- & \makecell[tl]{\citealp{comis_x-ray_2011} \\ \citealp{tiwari_hercules_2021}}\\
MACS2228 & I & (337.1386, 20.6207) & 10.5 & 2.15 & 469 & 146 & $2.01\pm0.22$ & -- & \makecell[tl]{\citealp{botteon_planck_2022} \\ \citealp{comis_x-ray_2011} \\ \citealp{jia_xmm-newton_2008}}\\

%% file: tables/clumpings.tex
Cluster & $k_\text{min}^{-1}$ & $k_\text{max}^{-1}$ & $\alpha$ & $\left(\frac{\delta \rho}{\rho} \right)_{R_\text{2500c}}$ & $C_\text{L}$ & $C_\text{U}$ & $V_\text{1D,L}$ & $V_\text{1D,U}$ & $\left( P_K/P \right)_\text{L}$ & $\left( P_K/P \right)_\text{U}$\\ 
 & [kpc] & [kpc] &  & [$10^{-2}$] &  &  & [km/s] & [km/s] & [\%] & [\%]\\ 
\hline \\ 
\endfirsthead 
Cluster & $k_\text{min}^{-1}$ & $k_\text{max}^{-1}$ & $\alpha$ & $\left(\frac{\delta \rho}{\rho} \right)_{R_\text{2500c}}$ & $C_\text{L}$ & $C_\text{U}$ & $V_\text{1D,L}$ & $V_\text{1D,U}$ & $\left( P_K/P \right)_\text{L}$ & $\left( P_K/P \right)_\text{U}$\\ 
 & [kpc] & [kpc] &  & [$10^{-2}$] &  &  & [km/s] & [km/s] & [\%] & [\%]\\ 
\hline \\ 
\endhead 
CygnusA & 448 & 70 & ${0.39}^{+0.05}_{-0.05}$ & ${17.4}^{+5.4}_{-3.9}$ & ${1.031}^{+0.007}_{-0.007}$ & ${1.040}^{+0.002}_{-0.002}$ & ${242}^{+26}_{-28}$ & ${274}^{+8}_{-6}$ & ${4.5}^{+1.0}_{-1.0}$ & ${5.7}^{+0.3}_{-0.2}$\\
A1795 & 427 & 140 & ${0.37}^{+0.09}_{-0.10}$ & ${18.4}^{+12.7}_{-7.5}$ & ${1.020}^{+0.005}_{-0.005}$ & ${1.049}^{+0.005}_{-0.002}$ & ${181}^{+21}_{-23}$ & ${279}^{+14}_{-7}$ & ${3.1}^{+0.7}_{-0.7}$ & ${7.0}^{+0.7}_{-0.3}$\\
A2744 & 649 & 112 & ${0.14}^{+0.06}_{-0.06}$ & ${17.7}^{+7.0}_{-5.1}$ & ${1.044}^{+0.015}_{-0.016}$ & ${1.115}^{+0.069}_{-0.026}$ & ${288}^{+47}_{-57}$ & ${465}^{+123}_{-57}$ & ${5.2}^{+1.7}_{-1.8}$ & ${12.4}^{+6.1}_{-2.6}$\\
A2052 & 165 & 75 & ${0.11}^{+0.08}_{-0.07}$ & ${12.7}^{+4.8}_{-4.2}$ & ${1.010}^{+0.003}_{-0.003}$ & ${1.073}^{+0.091}_{-0.024}$ & ${88}^{+10}_{-14}$ & ${234}^{+116}_{-41}$ & ${1.5}^{+0.4}_{-0.5}$ & ${10.1}^{+10.0}_{-3.0}$\\
A2597 & 312 & 63 & ${0.37}^{+0.06}_{-0.06}$ & ${15.0}^{+5.7}_{-4.1}$ & ${1.018}^{+0.005}_{-0.005}$ & ${1.031}^{+0.002}_{-0.002}$ & ${143}^{+19}_{-22}$ & ${186}^{+7}_{-5}$ & ${3.1}^{+0.8}_{-0.9}$ & ${5.1}^{+0.4}_{-0.3}$\\
Phoenix & 323 & 189 & ${0.05}^{+0.08}_{-0.04}$ & ${13.5}^{+3.3}_{-5.1}$ & ${1.009}^{+0.005}_{-0.005}$ & ${1.168}^{+0.441}_{-0.093}$ & ${182}^{+42}_{-61}$ & ${788}^{+713}_{-264}$ & ${1.5}^{+0.8}_{-0.8}$ & ${22.5}^{+28.8}_{-11.1}$\\
Bullet & 535 & 88 & ${0.34}^{+0.05}_{-0.06}$ & ${28.0}^{+9.9}_{-7.2}$ & ${1.069}^{+0.016}_{-0.016}$ & ${1.116}^{+0.007}_{-0.005}$ & ${360}^{+39}_{-44}$ & ${468}^{+13}_{-9}$ & ${7.8}^{+1.6}_{-1.7}$ & ${12.6}^{+0.6}_{-0.4}$\\
A520 & 300 & 91 & ${0.26}^{+0.07}_{-0.07}$ & ${27.4}^{+11.7}_{-8.3}$ & ${1.053}^{+0.010}_{-0.010}$ & ${1.143}^{+0.025}_{-0.013}$ & ${300}^{+27}_{-28}$ & ${493}^{+42}_{-22}$ & ${6.1}^{+1.1}_{-1.1}$ & ${15.1}^{+2.2}_{-1.2}$\\
MACS0947 & 478 & 318 & ${0.08}^{+0.12}_{-0.06}$ & ${11.9}^{+5.4}_{-6.2}$ & ${1.005}^{+0.003}_{-0.003}$ & ${1.086}^{+0.234}_{-0.048}$ & ${114}^{+26}_{-33}$ & ${454}^{+420}_{-151}$ & ${0.9}^{+0.5}_{-0.5}$ & ${13.0}^{+22.7}_{-6.8}$\\
A2219 & 373 & 68 & ${0.23}^{+0.07}_{-0.07}$ & ${10.3}^{+4.5}_{-3.1}$ & ${1.012}^{+0.004}_{-0.004}$ & ${1.023}^{+0.006}_{-0.003}$ & ${191}^{+32}_{-36}$ & ${267}^{+33}_{-18}$ & ${1.8}^{+0.6}_{-0.6}$ & ${3.5}^{+0.9}_{-0.4}$\\
A1367 & 236 & 54 & ${0.14}^{+0.07}_{-0.07}$ & ${16.0}^{+6.1}_{-4.4}$ & ${1.029}^{+0.007}_{-0.008}$ & ${1.093}^{+0.058}_{-0.019}$ & ${135}^{+15}_{-19}$ & ${241}^{+66}_{-27}$ & ${3.5}^{+0.8}_{-0.9}$ & ${10.3}^{+5.4}_{-2.0}$\\
A3667 & 270 & 60 & ${0.28}^{+0.05}_{-0.05}$ & ${18.5}^{+5.0}_{-4.0}$ & ${1.029}^{+0.005}_{-0.006}$ & ${1.062}^{+0.006}_{-0.004}$ & ${195}^{+17}_{-21}$ & ${286}^{+13}_{-8}$ & ${3.4}^{+0.6}_{-0.7}$ & ${7.1}^{+0.6}_{-0.4}$\\
A2319 & 538 & 91 & ${0.40}^{+0.06}_{-0.06}$ & ${17.0}^{+6.3}_{-4.6}$ & ${1.029}^{+0.007}_{-0.008}$ & ${1.036}^{+0.002}_{-0.002}$ & ${254}^{+31}_{-38}$ & ${285}^{+8}_{-7}$ & ${4.3}^{+1.0}_{-1.1}$ & ${5.3}^{+0.3}_{-0.2}$\\
ZW0008 & 306 & 92 & ${0.05}^{+0.05}_{-0.03}$ & ${12.8}^{+2.4}_{-3.0}$ & ${1.018}^{+0.006}_{-0.006}$ & ${1.179}^{+0.402}_{-0.089}$ & ${150}^{+23}_{-26}$ & ${469}^{+376}_{-136}$ & ${2.2}^{+0.7}_{-0.7}$ & ${18.2}^{+23.7}_{-8.1}$\\
A2142 & 360 & 91 & ${0.77}^{+0.08}_{-0.08}$ & ${30.4}^{+14.7}_{-10.0}$ & ${1.026}^{+0.006}_{-0.006}$ & ${1.060}^{+0.004}_{-0.004}$ & ${232}^{+26}_{-27}$ & ${351}^{+13}_{-11}$ & ${3.9}^{+0.9}_{-0.8}$ & ${8.5}^{+0.6}_{-0.5}$\\
A1240 & 394 & 181 & ${0.09}^{+0.08}_{-0.06}$ & ${22.7}^{+8.6}_{-8.2}$ & ${1.038}^{+0.012}_{-0.012}$ & ${1.301}^{+0.591}_{-0.137}$ & ${180}^{+27}_{-32}$ & ${506}^{+365}_{-132}$ & ${4.5}^{+1.4}_{-1.4}$ & ${27.2}^{+25.3}_{-10.2}$\\
MACS1149 & 229 & 119 & ${0.06}^{+0.08}_{-0.04}$ & ${19.4}^{+5.0}_{-6.5}$ & ${1.022}^{+0.007}_{-0.008}$ & ${1.311}^{+0.758}_{-0.159}$ & ${205}^{+32}_{-39}$ & ${767}^{+654}_{-230}$ & ${2.7}^{+0.9}_{-0.9}$ & ${27.8}^{+29.1}_{-11.9}$\\
MACS0416 & 408 & 121 & ${0.06}^{+0.07}_{-0.04}$ & ${18.0}^{+4.6}_{-5.9}$ & ${1.036}^{+0.016}_{-0.015}$ & ${1.285}^{+0.680}_{-0.146}$ & ${299}^{+59}_{-70}$ & ${838}^{+705}_{-253}$ & ${5.3}^{+2.1}_{-2.1}$ & ${30.5}^{+29.3}_{-12.9}$\\
A773 & 467 & 178 & ${0.06}^{+0.09}_{-0.05}$ & ${21.5}^{+6.5}_{-9.0}$ & ${1.043}^{+0.024}_{-0.024}$ & ${1.378}^{+0.992}_{-0.210}$ & ${250}^{+61}_{-84}$ & ${741}^{+670}_{-247}$ & ${5.1}^{+2.6}_{-2.8}$ & ${31.9}^{+31.0}_{-14.7}$\\
A780 & 158 & 74 & ${0.10}^{+0.09}_{-0.07}$ & ${18.7}^{+7.0}_{-6.5}$ & ${1.021}^{+0.006}_{-0.006}$ & ${1.170}^{+0.253}_{-0.058}$ & ${144}^{+21}_{-22}$ & ${413}^{+239}_{-78}$ & ${3.4}^{+1.0}_{-0.9}$ & ${22.8}^{+19.6}_{-6.5}$\\
A2034 & 325 & 82 & ${0.29}^{+0.07}_{-0.07}$ & ${18.1}^{+7.7}_{-5.5}$ & ${1.024}^{+0.007}_{-0.006}$ & ${1.057}^{+0.007}_{-0.004}$ & ${191}^{+25}_{-27}$ & ${294}^{+18}_{-11}$ & ${2.9}^{+0.8}_{-0.7}$ & ${6.6}^{+0.8}_{-0.5}$\\
A98N & 190 & 87 & ${0.18}^{+0.14}_{-0.11}$ & ${25.9}^{+19.2}_{-13.0}$ & ${1.040}^{+0.018}_{-0.018}$ & ${1.193}^{+0.244}_{-0.062}$ & ${167}^{+34}_{-44}$ & ${369}^{+186}_{-65}$ & ${5.8}^{+2.4}_{-2.6}$ & ${22.9}^{+17.3}_{-6.1}$\\
A1664 & 442 & 161 & ${0.45}^{+0.10}_{-0.10}$ & ${17.2}^{+12.8}_{-7.3}$ & ${1.018}^{+0.007}_{-0.007}$ & ${1.033}^{+0.005}_{-0.003}$ & ${136}^{+24}_{-29}$ & ${188}^{+12}_{-9}$ & ${2.6}^{+1.0}_{-1.0}$ & ${4.9}^{+0.6}_{-0.4}$\\
A1650 & 333 & 97 & ${0.73}^{+0.09}_{-0.10}$ & ${16.8}^{+11.9}_{-6.4}$ & ${1.012}^{+0.004}_{-0.004}$ & ${1.020}^{+0.001}_{-0.001}$ & ${134}^{+20}_{-21}$ & ${169}^{+4}_{-4}$ & ${1.9}^{+0.6}_{-0.5}$ & ${2.9}^{+0.1}_{-0.1}$\\
A133 & 98 & 62 & ${0.08}^{+0.12}_{-0.06}$ & ${11.8}^{+3.4}_{-4.7}$ & ${1.005}^{+0.003}_{-0.003}$ & ${1.091}^{+0.200}_{-0.039}$ & ${80}^{+18}_{-27}$ & ${336}^{+265}_{-83}$ & ${0.9}^{+0.4}_{-0.5}$ & ${13.6}^{+19.9}_{-5.4}$\\
MACS0717 & 444 & 99 & ${0.33}^{+0.08}_{-0.08}$ & ${32.9}^{+16.4}_{-11.0}$ & ${1.097}^{+0.027}_{-0.030}$ & ${1.165}^{+0.023}_{-0.014}$ & ${575}^{+75}_{-95}$ & ${749}^{+51}_{-32}$ & ${10.8}^{+2.6}_{-3.0}$ & ${17.0}^{+1.9}_{-1.2}$\\
3C186 & 287 & 228 & ${0.07}^{+0.11}_{-0.05}$ & ${19.8}^{+6.4}_{-9.2}$ & ${1.009}^{+0.005}_{-0.005}$ & ${1.298}^{+0.911}_{-0.181}$ & ${132}^{+36}_{-45}$ & ${769}^{+779}_{-288}$ & ${1.5}^{+0.9}_{-0.8}$ & ${34.1}^{+33.6}_{-17.2}$\\
ZW2341 & 407 & 126 & ${0.05}^{+0.07}_{-0.04}$ & ${17.3}^{+4.3}_{-5.8}$ & ${1.034}^{+0.022}_{-0.019}$ & ${1.274}^{+0.674}_{-0.148}$ & ${197}^{+57}_{-66}$ & ${557}^{+479}_{-180}$ & ${4.1}^{+2.5}_{-2.2}$ & ${25.4}^{+28.7}_{-11.9}$\\
MACS1347 & 459 & 164 & ${0.04}^{+0.05}_{-0.03}$ & ${22.8}^{+4.1}_{-6.1}$ & ${1.050}^{+0.014}_{-0.015}$ & ${1.651}^{+1.631}_{-0.343}$ & ${435}^{+59}_{-72}$ & ${1565}^{+1364}_{-489}$ & ${7.2}^{+1.9}_{-2.1}$ & ${50.1}^{+27.8}_{-17.9}$\\
A644 & 312 & 109 & ${0.20}^{+0.10}_{-0.10}$ & ${18.1}^{+11.5}_{-7.2}$ & ${1.024}^{+0.005}_{-0.005}$ & ${1.083}^{+0.048}_{-0.015}$ & ${214}^{+23}_{-23}$ & ${402}^{+103}_{-38}$ & ${3.5}^{+0.8}_{-0.7}$ & ${11.3}^{+5.4}_{-1.8}$\\
A3847 & 437 & 285 & ${0.05}^{+0.08}_{-0.04}$ & ${8.7}^{+2.2}_{-3.3}$ & ${1.003}^{+0.002}_{-0.002}$ & ${1.076}^{+0.224}_{-0.046}$ & ${56}^{+15}_{-23}$ & ${265}^{+262}_{-100}$ & ${0.6}^{+0.3}_{-0.4}$ & ${11.6}^{+22.6}_{-6.8}$\\
AS0295 & 315 & 139 & ${0.11}^{+0.08}_{-0.06}$ & ${22.1}^{+9.1}_{-7.5}$ & ${1.033}^{+0.008}_{-0.008}$ & ${1.228}^{+0.280}_{-0.077}$ & ${231}^{+27}_{-32}$ & ${605}^{+298}_{-113}$ & ${3.9}^{+0.9}_{-1.0}$ & ${22.0}^{+16.6}_{-6.3}$\\
A1758N & 317 & 138 & ${0.10}^{+0.09}_{-0.07}$ & ${14.0}^{+6.0}_{-5.4}$ & ${1.014}^{+0.005}_{-0.005}$ & ${1.100}^{+0.171}_{-0.040}$ & ${158}^{+26}_{-32}$ & ${430}^{+277}_{-96}$ & ${1.6}^{+0.6}_{-0.6}$ & ${11.0}^{+14.1}_{-4.1}$\\
A85 & 352 & 114 & ${0.18}^{+0.12}_{-0.10}$ & ${15.0}^{+11.1}_{-7.0}$ & ${1.020}^{+0.006}_{-0.005}$ & ${1.062}^{+0.062}_{-0.017}$ & ${167}^{+22}_{-22}$ & ${296}^{+122}_{-45}$ & ${3.0}^{+0.8}_{-0.7}$ & ${8.8}^{+7.3}_{-2.3}$\\
A3411 & 356 & 103 & ${0.19}^{+0.09}_{-0.09}$ & ${18.9}^{+11.6}_{-7.3}$ & ${1.032}^{+0.009}_{-0.008}$ & ${1.095}^{+0.059}_{-0.019}$ & ${205}^{+26}_{-29}$ & ${354}^{+96}_{-38}$ & ${3.8}^{+1.0}_{-1.0}$ & ${10.5}^{+5.5}_{-2.0}$\\
MACS1206 & 203 & 108 & ${0.09}^{+0.11}_{-0.06}$ & ${25.6}^{+9.6}_{-11.1}$ & ${1.033}^{+0.015}_{-0.015}$ & ${1.375}^{+0.782}_{-0.162}$ & ${328}^{+66}_{-84}$ & ${1111}^{+840}_{-273}$ & ${4.8}^{+2.0}_{-2.1}$ & ${36.6}^{+27.4}_{-11.9}$\\
A401 & 356 & 101 & ${0.39}^{+0.09}_{-0.09}$ & ${20.2}^{+12.7}_{-7.5}$ & ${1.022}^{+0.005}_{-0.005}$ & ${1.053}^{+0.003}_{-0.002}$ & ${204}^{+22}_{-25}$ & ${317}^{+10}_{-6}$ & ${3.2}^{+0.7}_{-0.7}$ & ${7.5}^{+0.4}_{-0.3}$\\
PKS0745 & 331 & 132 & ${0.26}^{+0.15}_{-0.13}$ & ${10.4}^{+10.5}_{-5.7}$ & ${1.006}^{+0.002}_{-0.002}$ & ${1.021}^{+0.011}_{-0.003}$ & ${111}^{+19}_{-24}$ & ${210}^{+50}_{-15}$ & ${0.9}^{+0.3}_{-0.3}$ & ${3.1}^{+1.6}_{-0.4}$\\
A1835 & 549 & 181 & ${0.39}^{+0.10}_{-0.10}$ & ${14.7}^{+11.2}_{-6.5}$ & ${1.015}^{+0.005}_{-0.005}$ & ${1.028}^{+0.005}_{-0.003}$ & ${179}^{+28}_{-32}$ & ${245}^{+20}_{-12}$ & ${2.2}^{+0.7}_{-0.7}$ & ${4.1}^{+0.7}_{-0.4}$\\
SPT2043 & 282 & 216 & ${0.07}^{+0.11}_{-0.05}$ & ${24.2}^{+7.6}_{-10.7}$ & ${1.015}^{+0.009}_{-0.009}$ & ${1.445}^{+1.234}_{-0.260}$ & ${160}^{+43}_{-57}$ & ${874}^{+824}_{-310}$ & ${2.5}^{+1.5}_{-1.5}$ & ${43.6}^{+30.9}_{-19.3}$\\
A1689 & 562 & 128 & ${0.24}^{+0.09}_{-0.09}$ & ${16.1}^{+10.5}_{-6.4}$ & ${1.025}^{+0.009}_{-0.009}$ & ${1.054}^{+0.020}_{-0.009}$ & ${232}^{+38}_{-45}$ & ${339}^{+59}_{-29}$ & ${3.7}^{+1.3}_{-1.3}$ & ${7.7}^{+2.6}_{-1.2}$\\
A2256 & 490 & 76 & ${0.37}^{+0.06}_{-0.06}$ & ${17.7}^{+6.8}_{-4.9}$ & ${1.029}^{+0.009}_{-0.009}$ & ${1.042}^{+0.003}_{-0.002}$ & ${204}^{+30}_{-35}$ & ${243}^{+8}_{-7}$ & ${3.5}^{+1.1}_{-1.1}$ & ${4.9}^{+0.3}_{-0.3}$\\
A586 & 203 & 108 & ${0.06}^{+0.08}_{-0.04}$ & ${8.7}^{+2.0}_{-2.9}$ & ${1.004}^{+0.003}_{-0.003}$ & ${1.069}^{+0.179}_{-0.038}$ & ${90}^{+25}_{-35}$ & ${352}^{+316}_{-115}$ & ${0.8}^{+0.5}_{-0.5}$ & ${10.7}^{+19.4}_{-5.5}$\\
ZW0104 & 264 & 119 & ${0.18}^{+0.10}_{-0.10}$ & ${16.1}^{+10.6}_{-6.6}$ & ${1.014}^{+0.005}_{-0.004}$ & ${1.072}^{+0.055}_{-0.016}$ & ${140}^{+21}_{-24}$ & ${320}^{+104}_{-37}$ & ${1.7}^{+0.5}_{-0.5}$ & ${8.2}^{+5.4}_{-1.7}$\\
RXC1504 & 488 & 155 & ${0.42}^{+0.15}_{-0.14}$ & ${17.6}^{+22.0}_{-9.9}$ & ${1.020}^{+0.007}_{-0.007}$ & ${1.037}^{+0.009}_{-0.004}$ & ${239}^{+38}_{-49}$ & ${326}^{+36}_{-17}$ & ${3.4}^{+1.1}_{-1.2}$ & ${6.1}^{+1.3}_{-0.6}$\\
MACS1621 & 571 & 498 & ${0.05}^{+0.09}_{-0.04}$ & ${6.7}^{+2.1}_{-3.1}$ & ${1.001}^{+0.001}_{-0.001}$ & ${1.042}^{+0.127}_{-0.027}$ & ${39}^{+14}_{-22}$ & ${307}^{+309}_{-123}$ & ${0.1}^{+0.1}_{-0.1}$ & ${6.8}^{+15.9}_{-4.2}$\\
MACS0025 & 377 & 185 & ${0.13}^{+0.13}_{-0.09}$ & ${17.8}^{+11.7}_{-9.2}$ & ${1.020}^{+0.009}_{-0.010}$ & ${1.122}^{+0.238}_{-0.054}$ & ${188}^{+39}_{-54}$ & ${468}^{+335}_{-118}$ & ${2.4}^{+1.0}_{-1.2}$ & ${13.2}^{+17.7}_{-5.3}$\\
3C438 & 312 & 109 & ${0.20}^{+0.07}_{-0.07}$ & ${21.1}^{+9.0}_{-6.2}$ & ${1.031}^{+0.009}_{-0.008}$ & ${1.110}^{+0.036}_{-0.017}$ & ${223}^{+30}_{-32}$ & ${421}^{+63}_{-35}$ & ${3.7}^{+1.0}_{-1.0}$ & ${12.0}^{+3.3}_{-1.7}$\\
A478 & 435 & 292 & ${0.07}^{+0.10}_{-0.05}$ & ${8.0}^{+2.7}_{-3.6}$ & ${1.002}^{+0.001}_{-0.001}$ & ${1.047}^{+0.126}_{-0.027}$ & ${72}^{+14}_{-23}$ & ${319}^{+290}_{-108}$ & ${0.4}^{+0.2}_{-0.2}$ & ${7.6}^{+15.5}_{-4.1}$\\
A4067 & 141 & 85 & ${0.08}^{+0.11}_{-0.06}$ & ${26.0}^{+8.1}_{-10.8}$ & ${1.028}^{+0.014}_{-0.014}$ & ${1.434}^{+0.944}_{-0.208}$ & ${136}^{+30}_{-38}$ & ${535}^{+419}_{-149}$ & ${3.4}^{+1.5}_{-1.6}$ & ${35.0}^{+28.1}_{-13.1}$\\
MACS1226 & 355 & 240 & ${0.06}^{+0.09}_{-0.04}$ & ${22.6}^{+6.4}_{-9.1}$ & ${1.020}^{+0.007}_{-0.008}$ & ${1.434}^{+1.189}_{-0.255}$ & ${159}^{+28}_{-36}$ & ${748}^{+699}_{-269}$ & ${2.9}^{+1.1}_{-1.2}$ & ${40.1}^{+31.4}_{-18.5}$\\
A1413 & 293 & 132 & ${0.09}^{+0.09}_{-0.06}$ & ${15.1}^{+5.8}_{-5.7}$ & ${1.016}^{+0.005}_{-0.005}$ & ${1.129}^{+0.237}_{-0.054}$ & ${176}^{+26}_{-31}$ & ${505}^{+346}_{-121}$ & ${2.6}^{+0.8}_{-0.8}$ & ${18.2}^{+20.5}_{-6.8}$\\
A3112 & 357 & 102 & ${0.32}^{+0.09}_{-0.09}$ & ${13.0}^{+8.4}_{-5.2}$ & ${1.014}^{+0.004}_{-0.005}$ & ${1.026}^{+0.007}_{-0.004}$ & ${141}^{+19}_{-25}$ & ${192}^{+24}_{-14}$ & ${2.4}^{+0.7}_{-0.8}$ & ${4.4}^{+1.1}_{-0.6}$\\
A665 & 281 & 101 & ${0.06}^{+0.08}_{-0.04}$ & ${10.8}^{+2.8}_{-3.8}$ & ${1.011}^{+0.004}_{-0.005}$ & ${1.095}^{+0.223}_{-0.049}$ & ${123}^{+23}_{-33}$ & ${367}^{+304}_{-111}$ & ${1.3}^{+0.5}_{-0.6}$ & ${10.6}^{+17.7}_{-5.1}$\\
A1351 & 478 & 230 & ${0.14}^{+0.12}_{-0.09}$ & ${11.5}^{+8.3}_{-5.8}$ & ${1.009}^{+0.004}_{-0.004}$ & ${1.046}^{+0.074}_{-0.018}$ & ${127}^{+24}_{-30}$ & ${292}^{+181}_{-67}$ & ${1.1}^{+0.4}_{-0.4}$ & ${5.3}^{+7.6}_{-2.1}$\\
A1914 & 444 & 93 & ${0.06}^{+0.05}_{-0.04}$ & ${25.0}^{+6.3}_{-6.4}$ & ${1.088}^{+0.026}_{-0.026}$ & ${1.503}^{+0.872}_{-0.201}$ & ${381}^{+52}_{-61}$ & ${915}^{+597}_{-206}$ & ${9.8}^{+2.5}_{-2.7}$ & ${38.4}^{+24.6}_{-11.2}$\\
A2495 & 217 & 128 & ${0.10}^{+0.13}_{-0.07}$ & ${9.7}^{+4.3}_{-4.6}$ & ${1.004}^{+0.002}_{-0.002}$ & ${1.050}^{+0.107}_{-0.023}$ & ${71}^{+15}_{-18}$ & ${242}^{+186}_{-65}$ & ${0.7}^{+0.3}_{-0.3}$ & ${8.0}^{+13.4}_{-3.6}$\\
A2626 & 234 & 62 & ${0.41}^{+0.10}_{-0.10}$ & ${16.4}^{+10.3}_{-6.3}$ & ${1.018}^{+0.006}_{-0.006}$ & ${1.033}^{+0.004}_{-0.002}$ & ${119}^{+18}_{-22}$ & ${163}^{+10}_{-6}$ & ${2.6}^{+0.8}_{-0.9}$ & ${4.9}^{+0.6}_{-0.3}$\\
A2443 & 236 & 92 & ${0.23}^{+0.14}_{-0.12}$ & ${24.2}^{+20.5}_{-12.0}$ & ${1.034}^{+0.012}_{-0.012}$ & ${1.128}^{+0.083}_{-0.022}$ & ${234}^{+38}_{-46}$ & ${454}^{+129}_{-40}$ & ${4.1}^{+1.3}_{-1.4}$ & ${13.7}^{+7.0}_{-2.0}$\\
MACS1423 & 568 & 372 & ${0.07}^{+0.10}_{-0.05}$ & ${17.0}^{+5.9}_{-7.8}$ & ${1.012}^{+0.005}_{-0.005}$ & ${1.219}^{+0.585}_{-0.127}$ & ${170}^{+29}_{-40}$ & ${722}^{+662}_{-254}$ & ${2.1}^{+0.8}_{-0.8}$ & ${27.5}^{+30.7}_{-13.8}$\\
ZW0949 & 385 & 180 & ${0.11}^{+0.12}_{-0.08}$ & ${12.8}^{+7.3}_{-6.4}$ & ${1.011}^{+0.004}_{-0.005}$ & ${1.072}^{+0.157}_{-0.034}$ & ${128}^{+22}_{-30}$ & ${323}^{+253}_{-88}$ & ${1.9}^{+0.7}_{-0.8}$ & ${11.1}^{+17.3}_{-4.9}$\\
AS1063 & 505 & 272 & ${0.18}^{+0.14}_{-0.12}$ & ${14.0}^{+13.9}_{-8.0}$ & ${1.010}^{+0.004}_{-0.004}$ & ${1.054}^{+0.073}_{-0.018}$ & ${204}^{+39}_{-45}$ & ${476}^{+256}_{-90}$ & ${1.5}^{+0.6}_{-0.6}$ & ${7.6}^{+8.7}_{-2.5}$\\
A2199 & 211 & 72 & ${0.13}^{+0.06}_{-0.06}$ & ${7.1}^{+2.4}_{-1.8}$ & ${1.004}^{+0.001}_{-0.001}$ & ${1.019}^{+0.013}_{-0.005}$ & ${68}^{+8}_{-9}$ & ${142}^{+41}_{-19}$ & ${0.7}^{+0.2}_{-0.2}$ & ${2.9}^{+1.9}_{-0.7}$\\
MACS2243 & 314 & 162 & ${0.06}^{+0.08}_{-0.05}$ & ${18.9}^{+5.4}_{-7.0}$ & ${1.021}^{+0.009}_{-0.009}$ & ${1.290}^{+0.700}_{-0.145}$ & ${166}^{+34}_{-42}$ & ${619}^{+525}_{-182}$ & ${2.5}^{+1.1}_{-1.1}$ & ${26.4}^{+28.7}_{-11.2}$\\
MACS1931 & 426 & 312 & ${0.06}^{+0.10}_{-0.05}$ & ${13.3}^{+4.2}_{-5.8}$ & ${1.005}^{+0.003}_{-0.003}$ & ${1.142}^{+0.388}_{-0.080}$ & ${110}^{+29}_{-41}$ & ${577}^{+538}_{-195}$ & ${0.8}^{+0.5}_{-0.5}$ & ${17.9}^{+27.0}_{-9.2}$\\
RXC1532 & 395 & 191 & ${0.19}^{+0.14}_{-0.12}$ & ${16.5}^{+15.2}_{-9.0}$ & ${1.016}^{+0.006}_{-0.007}$ & ${1.072}^{+0.090}_{-0.024}$ & ${163}^{+30}_{-43}$ & ${351}^{+176}_{-65}$ & ${2.6}^{+1.0}_{-1.2}$ & ${11.1}^{+10.8}_{-3.4}$\\
A2204 & 746 & 226 & ${0.33}^{+0.12}_{-0.12}$ & ${18.4}^{+18.8}_{-9.4}$ & ${1.025}^{+0.009}_{-0.009}$ & ${1.052}^{+0.016}_{-0.007}$ & ${186}^{+32}_{-35}$ & ${268}^{+39}_{-20}$ & ${3.7}^{+1.3}_{-1.2}$ & ${7.4}^{+2.1}_{-1.0}$\\
4C+37.11 & 312 & 91 & ${0.48}^{+0.09}_{-0.09}$ & ${18.6}^{+11.9}_{-7.3}$ & ${1.021}^{+0.007}_{-0.007}$ & ${1.037}^{+0.003}_{-0.002}$ & ${174}^{+26}_{-31}$ & ${231}^{+10}_{-7}$ & ${3.5}^{+1.1}_{-1.1}$ & ${6.0}^{+0.5}_{-0.3}$\\
RXC0528 & 431 & 200 & ${0.31}^{+0.15}_{-0.15}$ & ${28.9}^{+37.0}_{-16.6}$ & ${1.046}^{+0.014}_{-0.013}$ & ${1.136}^{+0.080}_{-0.026}$ & ${314}^{+44}_{-47}$ & ${539}^{+141}_{-54}$ & ${6.6}^{+1.8}_{-1.8}$ & ${17.3}^{+7.7}_{-2.8}$\\
ZW1454 & 310 & 183 & ${0.06}^{+0.08}_{-0.04}$ & ${17.7}^{+4.5}_{-6.5}$ & ${1.016}^{+0.008}_{-0.008}$ & ${1.285}^{+0.731}_{-0.159}$ & ${135}^{+30}_{-40}$ & ${578}^{+514}_{-195}$ & ${2.3}^{+1.1}_{-1.2}$ & ${30.5}^{+30.5}_{-14.3}$\\
A2029 & 312 & 94 & ${0.65}^{+0.12}_{-0.13}$ & ${12.8}^{+12.1}_{-5.8}$ & ${1.006}^{+0.002}_{-0.002}$ & ${1.013}^{+0.001}_{-0.001}$ & ${113}^{+16}_{-19}$ & ${166}^{+7}_{-5}$ & ${0.9}^{+0.3}_{-0.3}$ & ${1.9}^{+0.2}_{-0.1}$\\
A1300 & 340 & 200 & ${0.05}^{+0.06}_{-0.03}$ & ${28.5}^{+5.9}_{-8.4}$ & ${1.040}^{+0.010}_{-0.009}$ & ${1.893}^{+2.334}_{-0.484}$ & ${283}^{+33}_{-35}$ & ${1331}^{+1199}_{-430}$ & ${4.8}^{+1.1}_{-1.1}$ & ${52.5}^{+27.5}_{-18.9}$\\
A2390 & 433 & 124 & ${0.07}^{+0.07}_{-0.05}$ & ${15.5}^{+4.5}_{-4.7}$ & ${1.026}^{+0.007}_{-0.007}$ & ${1.173}^{+0.302}_{-0.071}$ & ${275}^{+35}_{-43}$ & ${709}^{+466}_{-164}$ & ${3.8}^{+1.0}_{-1.1}$ & ${21.0}^{+21.2}_{-7.4}$\\
RXC0439 & 490 & 333 & ${0.07}^{+0.11}_{-0.05}$ & ${11.9}^{+4.4}_{-5.7}$ & ${1.005}^{+0.004}_{-0.004}$ & ${1.101}^{+0.302}_{-0.060}$ & ${97}^{+28}_{-41}$ & ${417}^{+415}_{-152}$ & ${0.9}^{+0.6}_{-0.6}$ & ${15.0}^{+26.2}_{-8.3}$\\
A2537 & 368 & 251 & ${0.06}^{+0.09}_{-0.05}$ & ${9.0}^{+2.6}_{-3.7}$ & ${1.003}^{+0.002}_{-0.002}$ & ${1.066}^{+0.185}_{-0.039}$ & ${82}^{+20}_{-28}$ & ${393}^{+371}_{-140}$ & ${0.5}^{+0.3}_{-0.3}$ & ${10.3}^{+20.0}_{-5.8}$\\
MACS2140 & 606 & 446 & ${0.06}^{+0.10}_{-0.05}$ & ${8.2}^{+2.7}_{-3.7}$ & ${1.002}^{+0.001}_{-0.001}$ & ${1.055}^{+0.163}_{-0.033}$ & ${73}^{+22}_{-33}$ & ${382}^{+377}_{-142}$ & ${0.4}^{+0.2}_{-0.2}$ & ${8.7}^{+18.7}_{-5.1}$\\
ZW1518 & 260 & 120 & ${0.11}^{+0.09}_{-0.07}$ & ${8.4}^{+3.7}_{-3.1}$ & ${1.005}^{+0.001}_{-0.001}$ & ${1.033}^{+0.053}_{-0.013}$ & ${71}^{+9}_{-11}$ & ${187}^{+114}_{-40}$ & ${0.8}^{+0.2}_{-0.2}$ & ${5.4}^{+7.5}_{-2.0}$\\
A2063 & 358 & 145 & ${0.41}^{+0.16}_{-0.16}$ & ${17.2}^{+24.3}_{-10.0}$ & ${1.018}^{+0.005}_{-0.005}$ & ${1.036}^{+0.013}_{-0.005}$ & ${138}^{+19}_{-21}$ & ${195}^{+32}_{-13}$ & ${3.0}^{+0.8}_{-0.8}$ & ${5.9}^{+1.9}_{-0.7}$\\
A2151 & 196 & 67 & ${0.18}^{+0.19}_{-0.12}$ & ${19.4}^{+15.6}_{-11.5}$ & ${1.031}^{+0.013}_{-0.012}$ & ${1.105}^{+0.176}_{-0.038}$ & ${127}^{+24}_{-26}$ & ${233}^{+148}_{-47}$ & ${4.6}^{+1.8}_{-1.7}$ & ${13.9}^{+16.3}_{-4.6}$\\
MACS2228 & 407 & 279 & ${0.08}^{+0.12}_{-0.06}$ & ${21.2}^{+9.1}_{-10.8}$ & ${1.016}^{+0.009}_{-0.009}$ & ${1.273}^{+0.725}_{-0.154}$ & ${199}^{+49}_{-63}$ & ${821}^{+748}_{-278}$ & ${2.4}^{+1.3}_{-1.3}$ & ${29.6}^{+31.0}_{-14.1}$